\documentclass[a4paper,fleqn,usenatbib]{mnras}

\usepackage[T1]{fontenc}
\usepackage{ae,aecompl}
\usepackage{graphicx}	
\usepackage{amsmath}	
\usepackage{amssymb}	

\usepackage{color}
\usepackage{rotating,lscape}

\newcommand{\HB}{H{$\beta$}}
\newcommand{\Halpha}{H$\alpha$}
\def\p0{\phantom{0}}
\def\HII{\hbox{H\,{\sc ii}}}
\def\D{\hbox{$^\circ$}}

\title[RC PNe in the LMC]{Radio Planetary Nebulae in the Large Magellanic Cloud}

\author[H. Leverenz et al.]{
Howard Leverenz,$^{1,2}$\thanks{E-mail: hleverenz@sbcglobal.net}
Miroslav D. Filipovi\'c,$^{1}$
B. Vukoti{\'c},$^{3}$
D. Uro{\v s}evi{\'c},$^{4,5}$
\newauthor
Kevin Grieve$^{1}$
\\
$^{1}$Western Sydney University, Locked Bag 1797, Penrith, NSW 2751, Australia\\
$^{2}$University of Southern Queensland, Toowoomba, Queensland 4350, Australia\\
$^{3}$Astronomical Observatory, Volgina 7, 11060 Belgrade 38, Serbia\\
$^{4}$Department of Astronomy, Faculty of Mathematics, University of Belgrade, Studentski trg 16, 11000 Belgrade, Serbia\\
$^{5}$Isaac Newton Institute of Chile, Yugoslavia Branch\\
}

\date{Accepted XXX. Received YYY; in original form ZZZ}

\pubyear{2017}

\begin{document}
\label{firstpage}
\pagerange{\pageref{firstpage}--\pageref{lastpage}}
\maketitle


\begin{abstract}
{We present 21 new radio-continuum detections at catalogued planetary nebula (PN) positions in the Large Magellanic Cloud (LMC) using all presently available data from the Australia Telescope Online Archive at 3, 6, 13 and 20~cm. Additionally, 11 previously detected LMC radio PNe are re-examined with $ 7 $ detections confirmed and reported here. An additional three PNe from our previous surveys are also studied. The last of the 11 previous detections is now classified as a compact \HII\ region which makes for a total sample of 31 radio PNe in the LMC. The radio-surface brightness to diameter ($\Sigma$--D) relation is parametrised as $\Sigma \propto {D^{ - \beta }}$. With the available 6~cm $\Sigma$--$D$ data we construct $\Sigma$--$D$ samples from 28 LMC PNe and 9 Small Magellanic Cloud (SMC) radio detected PNe. The results of our sampled PNe in the Magellanic Clouds (MCs) are comparable to previous measurements of the Galactic PNe. We obtain $\beta=2.9\pm0.4$ for the MC PNe compared to $\beta = 3.1\pm0.4$ for the Galaxy. For a better insight into sample completeness and evolutionary features we reconstruct the $\Sigma$--$D$ data probability density function (PDF). The PDF analysis implies that PNe are not likely to follow linear evolutionary paths. To estimate the significance of sensitivity selection effects we perform a Monte Carlo sensitivity simulation on the $\Sigma$--$D$ data. The results suggest that selection effects are significant for values larger than $\beta \sim 2.6$ and that a measured slope of $\beta=2.9$ should correspond to a sensitivity-free value of $\sim 3.4$. }

\end{abstract}




\section{Introduction}
 \label{s:intro}

\citet{Filipovic2009} reported the first confirmed extragalactic radio-continuum (RC) detection of 15 planetary nebulae (PNe) in the Magellanic Clouds (MCs) of which 11 are located in the Large Magellanic Cloud (LMC). Prior to this study, a tentative radio detection of only three extragalactic PNe had been reported in the literature \citep{1994A&A...290..228Z,2000A&A...363..717D}. \cite{Bojicic2010} and \cite{leverenz2016}, reported detection results and RC data measurements from 16 PN positions in the Small Magellanic Cloud (SMC). The RC detection of MC PNe is re-examined and extended  in the present paper with a total of 31 RC detections. These consist of 21 new RC detections reported here for the first time, re-measurements of seven previously reported RC detections in the LMC, and three PNe which were detected in  LMC surveys and reported in \citet{Filipovic2009}.

\section{Archival data and new radio-continuum observations}
 \label{s:data}

The Australia Telescope Compact Array (ATCA) archive databases contain unprocessed Australia Telescope National Facility (ATNF) radio astronomy data obtained with the ATCA and other Australian facilities managed by the Commonwealth Scientific and Industrial Research Organisation of Australia (CSIRO). This data is accessible through the Australia Telescope Online Archive (ATOA) search page\footnote{http://atoa.atnf.csiro.au/}. We first searched the ATOA to select any projects with data at 3, 6, 13 or 20~cm that were obtained using a 6~km configuration of the ATCA antenna array in order to select projects with the best possible resolution. Typically, these projects contained raw data for multiple images with different central coordinates. A project was selected for processing if an LMC PN was located within a 10~arcmin radius of any central coordinate contained therein. The base catalogue and the criteria for selection are discussed in Sect.~\ref{s:RC Detection}. Most of these ATCA projects were intended to image various other LMC objects that were not PNe.

This manual search identified projects that could possibly contain PNe in images that might result in a PN RC detection. Each central coordinate and each frequency from all of the identified projects found in this way were manually processed to create images from the data and then the images were examined at the coordinates of known (optical) PNe for RC emission detection. 

Our project, ATCA C2367, which operated in 2010, was executed specifically to study RC PNe in the MCs. This project yielded most (26 out of 29) of the 3~cm and 6~cm detections in this study.

Seventeen ATCA projects were found to meet our selection criteria as explained above. All of the raw data at our frequencies of interest from these projects were examined. Each project contained data from two to four frequencies. We manually processed all of the data to produce images at every frequency. Every image was examined for RC detections at the PN coordinates. Due to differences in sensitivity (RMS noise level) and primary beam coverage from one image/frequency to another, PN RC detection at one frequency did not mean that detection at another frequency was to be expected. Only nine projects (see Table~\ref{tbl:data}) had data that led to a PN RC detection. Four of those projects had PN RC detections at only one frequency. 

Table~\ref{tbl:ProjNoFind} lists the projects that had images that did not contain PN detections and shows the pointing centres of each of those images. It also contains the RP catalogue references (see Sect.~\ref{s:RC Detection}) to the PNe contained within approximately two thirds of the diameter of each full image since detections close to the edge of the images are not reliable. A total of 61 images were developed that did not contain any PN RC detections.

The ATCA data were processed using {\sc miriad} software \citep{2011ascl.soft06007S}. The {\sc miriad~\sc sigest} task was used to determine the 1$\sigma$ RMS noise level in a region essentially devoid of any strong point sources near each PN examined. For unresolved sources, the {\sc miriad~\sc imfit} task was used to determine the integrated flux density from the PN using a Gaussian-fitting. For resolved sources, the pixels within a 1$\sigma$ Jy~beam$^{-1}$ contour centred on the PN were used to estimate the PN integrated flux density. A positive RC detection with either method was defined as a flux density estimate $\geqslant$3$\sigma$ above the RMS noise.

%
 \begin{table}
 \centering
 \normalsize
 \caption{ {Radio data used in this study are identified with the ATCA project numbers. Data using the Compact Array Broadband Backend (CABB) configuration were acquired using the 2~GHz receiver bandwidth capability of the ATCA. Project C2908 is a MOSAIC survey of the LMC Supershell~4. } } 
 \label{tbl:data}
 \begin{tabular}{lcccc}
 \hline  
 ATCA           & $\nu$		& $\lambda$ & Beam          & RMS\\
 Project No.    & (GHz)     & (cm) 	    & ($\arcsec \times \arcsec$)      & (mJy/beam) \\
 \hline
 C256	    	& 1.376     & 20  	&  2.4$\times$7.0 	& 0.1\p0\\
 C256	    	& 2.378     & 13   	&  7.7$\times$4.2 	& 0.1\p0\\
 C308	    	& 5.824     & \p06	&  2.5$\times$1.1 	& 0.1\p0\\
 C308	    	& 8.640     & \p03  	&  1.6$\times$0.7  	& 0.1\p0\\
 C354	    	& 1.376     & 20  	&  8.5$\times$7.9  	& 0.15\\
 C395		    & 1.376     & 20  	&  6.6$\times$6.3  	& 0.13\\
 C395		    & 2.378     & 13  	&  3.8$\times$3.5  	& 0.14\\
 C479		    & 1.344     & 20  	&  7.4$\times$6.9  	& 0.12\\
 C479		    & 2.378     & 13  	&  4.1$\times$3.9  	& 0.12\\
 C520		    & 1.384     & 20 	&  10.0$\times$6.0 	& 0.15\\
 C1973 CABB	    & 5.500     & \p06	&  2.4$\times$2.0  	& 0.03\\
 C2367 CABB	    & 5.500     & \p06	&  2.0$\times$1.6  	& 0.01\\
 C2367 CABB	    & 9.000     & \p03  	&  1.2$\times$1.0  	& 0.1\p0\\
 C2908 CABB	    & 2.162     & 13		&  5.3$\times$3.9  	& 0.3\p0\\
 \hline
 \end{tabular}
 \end{table}

\begin{table} 
\centering
\scriptsize
\caption{ {ATCA projects that contained PN coordinates but did not yield PN RC detections. The listed coordinates represent the centre of the pointing.}}   
 \begin{tabular}{lcccl}
 \hline
ATCA 	&$\lambda$ 	& RA(J2000)	& DEC (J2000)					&RP Catalogued PNe*  \\
Project	&(cm)	    & (h m s)	& (\D~\arcmin~\arcsec)			&not  detected \\
\hline
C015	&6		    &05:35:27.90&--69:16:21.59  	& 365 395 \\
\smallskip
		&3			&--			&--			 	& 395 \\
C148	&6			&05:08:59.00&--68:43:34.09	& 1395 1396 1399 1426 \\
        &       	&           &               & 1443 1444 1446 1456 \\
        &       	&           &               & 1550 \\
\smallskip
		&3			&--			&--				& 1444 1550 \\
C256	&20			&05:18:51.40&--69:37:31.70 	& 1223 1225 1228 1229 \\
		&       	&           &               & 1230 1231 1232 1234 \\
		&       	&           &               & 1235 1338 1341 1352\\
		&		& 			& 				&  1353 1354 1357 1358\\
		&       	&           &               &  1371 2194\\
		&13			&--			&--				& 1225 1228 1229 1230 \\
\smallskip
		&       	&           &               & 1338 1341 \\
C282	&20		    &04:53:29.70&--67:23:20.20	& 1078 1081 1092 1095 \\
		&       	&           &               & 1554 1555 1556 1558 \\
		&       	&           &               & 1580 1584 1595  \\
		&13 		&--			&--				& 1092 1095 \\
		&20	    	&05:22:41.10&--66:40:56.30	& 1801 1895 \\
\smallskip
		&13			&--			&--				& 1895 \\
C354	&20			&05:09:53.00&--68:53:15.00	& 1218 1219 1310 1313 \\
		&       	&           &               & 1314 1315 1317 1323 \\
		&       	&           &               & 1324 1395 1396 1399 \\
		&       	&           &               & 1400 1402 1408 1416 \\
		&			&			&  				& 1418 1426 1443 1444 \\
		&       	&           &               & 1446 1456 1550 1823 \\
		&       	&           &               & 1835 \\
		&13			&--			&--				& 1218 1323 1324 1395 \\
		&       	&           &               & 1399 1418 1426 1443 \\
		&       	&           &               & 1444 1446 1550\\
		&20			&05:35:24.00&--67:34:50.00	& 366 1053 \\
\smallskip
		&13		&--				&--				& 366 1053 \\
C394	&20		&05:05:54.00   &--68:01:46.00	& 1397 1409 1532 1798 \\
		&			&			&  				& 1802 1878 \\
\smallskip
		&13		&--				&--				& 1802 1878  \\

C395	&20		&04:59:42.00   &--70:09:10.00	& 1602 1605 1608 1609 \\
\smallskip
		&			&			&  				& 1660 1679 1697 1771 \\
C479	&20		&05:05:30.00   &--67:52:00.00	& 1397 1532 1802 1878 \\
		&			&			&  				& 1886 \\
\smallskip
		&13		&--				&--				& 1878 \\
\smallskip
C520	&20		&05:13:50.80   &--69:51:47.00	& 1225 1237 1267 1288 \\

C634	&20		&05:13:50.80   &--69:51:47.00	& 1636 \\
\smallskip
		&13		&--				&--				& 1636 \\
\smallskip
C663			&6		&05:31:55.82   &--71:00:18.86	& 1012 \\
C1074	&20		&04:53:40.00   &--68:29:40.00	& 1800 \\
\smallskip
		&13		&--				&--				& 1800 \\
C2044	&6		&05:23:33.02   &--69:37:03.73	& 757 875 889 \\
		&3		&--				&--				& 757 875 889  \\
		&6		&05:27:41.04   &--67:27:15.60	& 1047 \\
		&6		&05:39:07.52   &--69:30:25.44	& 243 \\
		&6		&05:39:57.98   &--71:10:18.05	& 44 46 \\
		&3		&--				&--				& 44 46 \\
		&6		&05:19:11.78   &--69:12:23.84	& 1227 \\
		&3		&--				&--				& 1227 \\
		&6		&05:22:08.68   &--67:58:12.87	& 979 980 1534 \\
\smallskip
		&3		&--				&--				& 980 979 \\
C2367	&6		&05:03:41.30   &--70:14:13.60	& 1605 \\
		&3		&--				&--				& 1605 \\
\hline
\end{tabular}
\begin{flushleft}
RP Catalogued PNe* This catalogue is a general inventory of LMC PNe known at time of publication and also contains PNe found by  \cite{Henize1956}, \cite{Lindsay1961}, \cite{Lindsay1963}, \cite{Lindsay1963a}, \cite{Westerlund1964}, \cite{Sanduleak1978}, \cite{Jacoby1980}, \cite{Morgan1992}, \cite{Morgan1994}.
\end{flushleft}
\label{tbl:ProjNoFind}
\end{table}

\section{Method and Results}
 \label{s:RC Detection}

\subsection{Base catalogue and detection method}

In order to make the base list for this study we used the PN catalogue from \citet{Reid2006a,Reid2006,Reid2010}, collectively referred to here as RP. The RP catalogue of the LMC was constructed from a portion of the Anglo-Australian Observatory (AAO)/UKST \Halpha\ survey of the Southern Galactic Plane \citep{Parker2005}. The catalogue covers $\approx 25~deg^2$ centred on the LMC and contains 629 PNe found by using deep \Halpha\ and `Short Red' images. Each PN was then verified using spectroscopic measurements and classified as `Known', `True', `Likely', or `Possible'. This catalogue was further refined with an extensive re-examination of the 'Likely' and 'Possible' categories using a multiwavelength analysis of the catalogue in \cite{Reid2014}. The subset of the RP catalogue used in our current study contains coordinates only from PNe classified as `Known' or `True'. 

The RP catalogue numbers are written as `RPxxxx' to stand for `[RP2006]~xxxx'. LMC catalogue references from \cite{Sanduleak1978}, which have the form `SMP\,LMC\,xx', are referred to as `SMP\,Lxx'. For references to SMC PNe catalogued as `SMP\,SMC\,xx', we use `SMP\,Sxx' as a shorthand. In general, we use SMP catalogue numbers where possible, but for PNe which do not appear in the SMP catalogue, we use the RP catalogue numbers. 

\begin{figure*}
\centering
\includegraphics[angle=-90, trim=0 0 0 0, width=.475\textwidth]{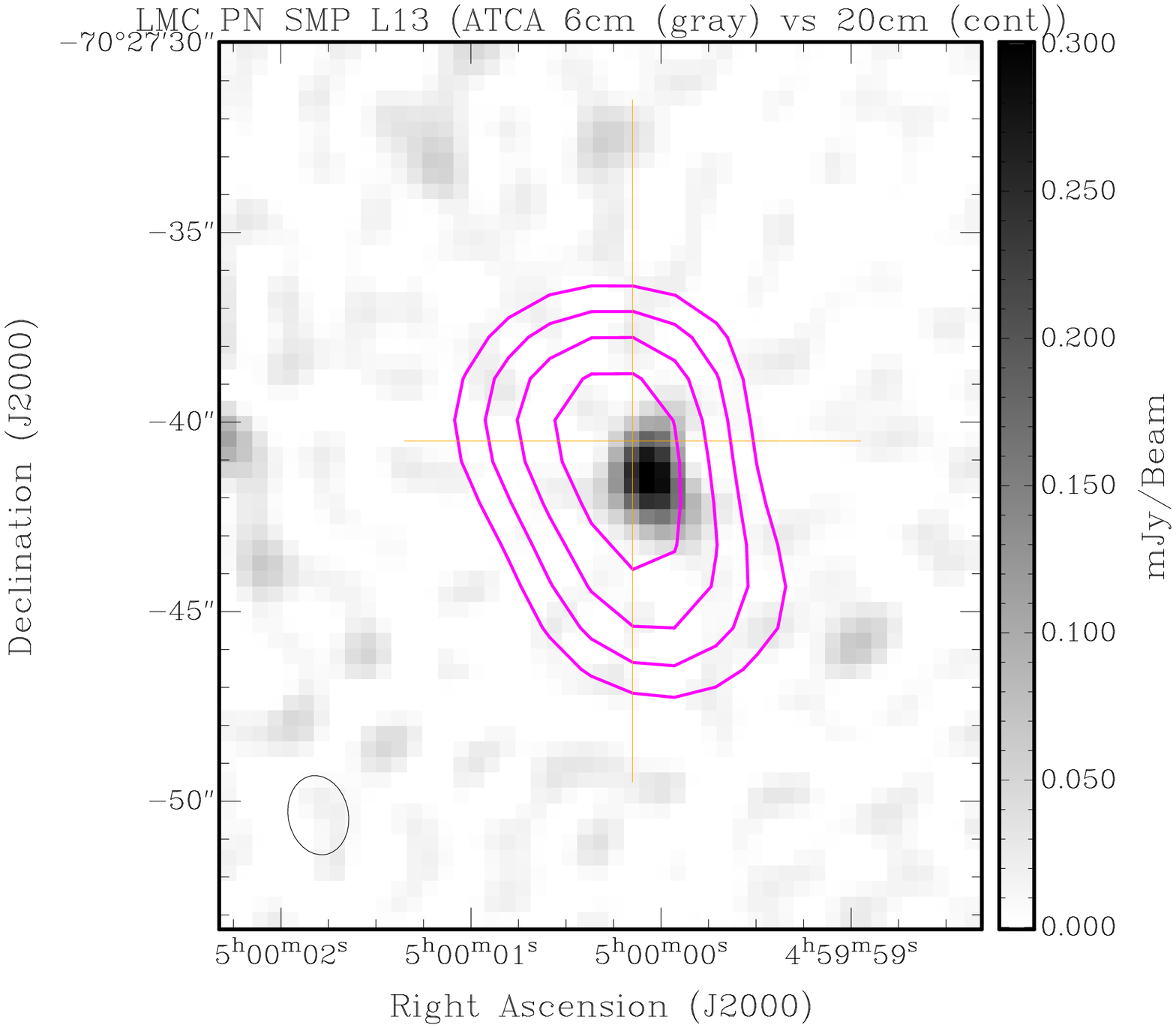}
\includegraphics[angle=-90, trim=0 0 0 0, width=.475\textwidth]{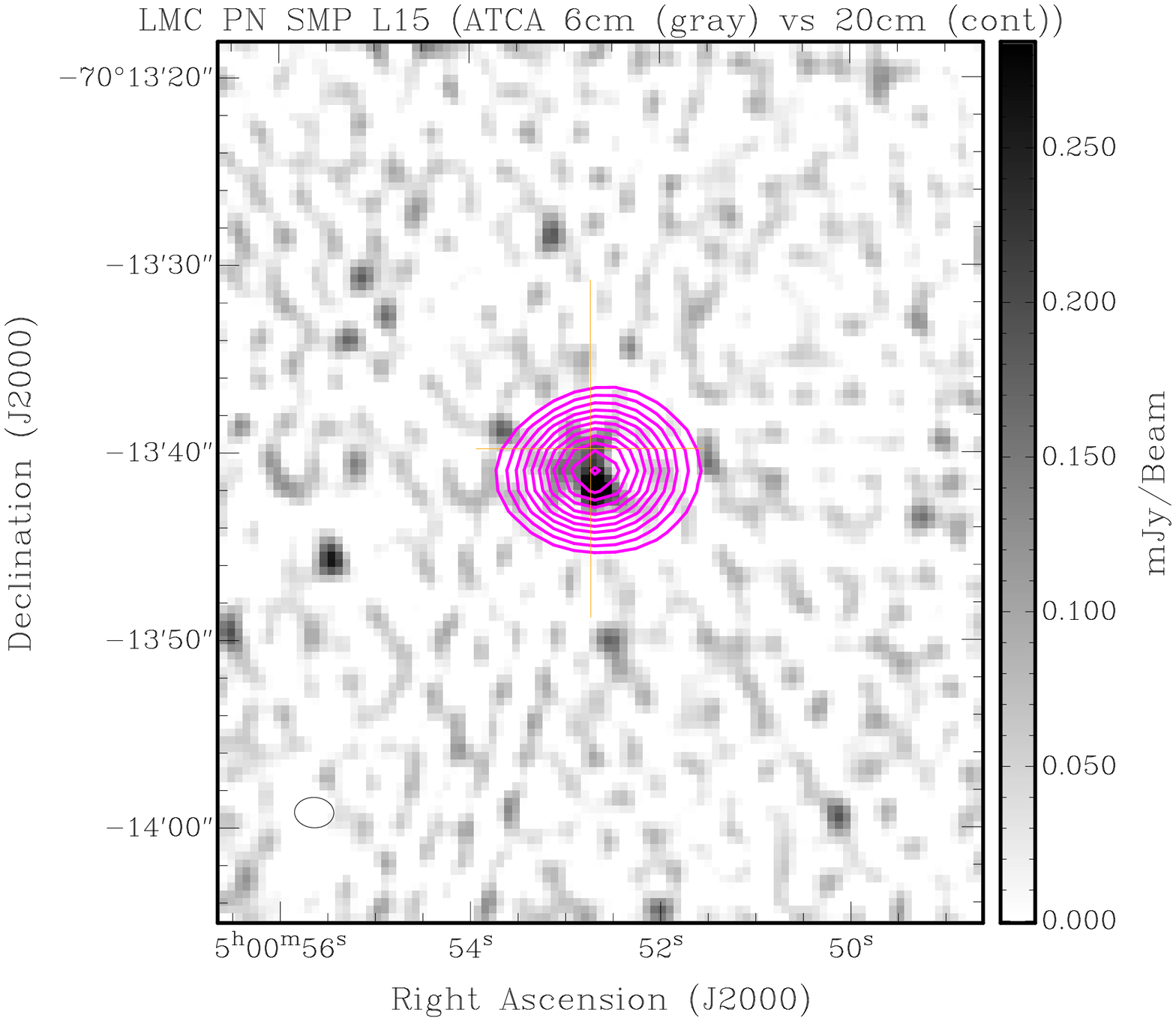}
\includegraphics[angle=-90, trim=0 0 0 0, width=.475\textwidth]{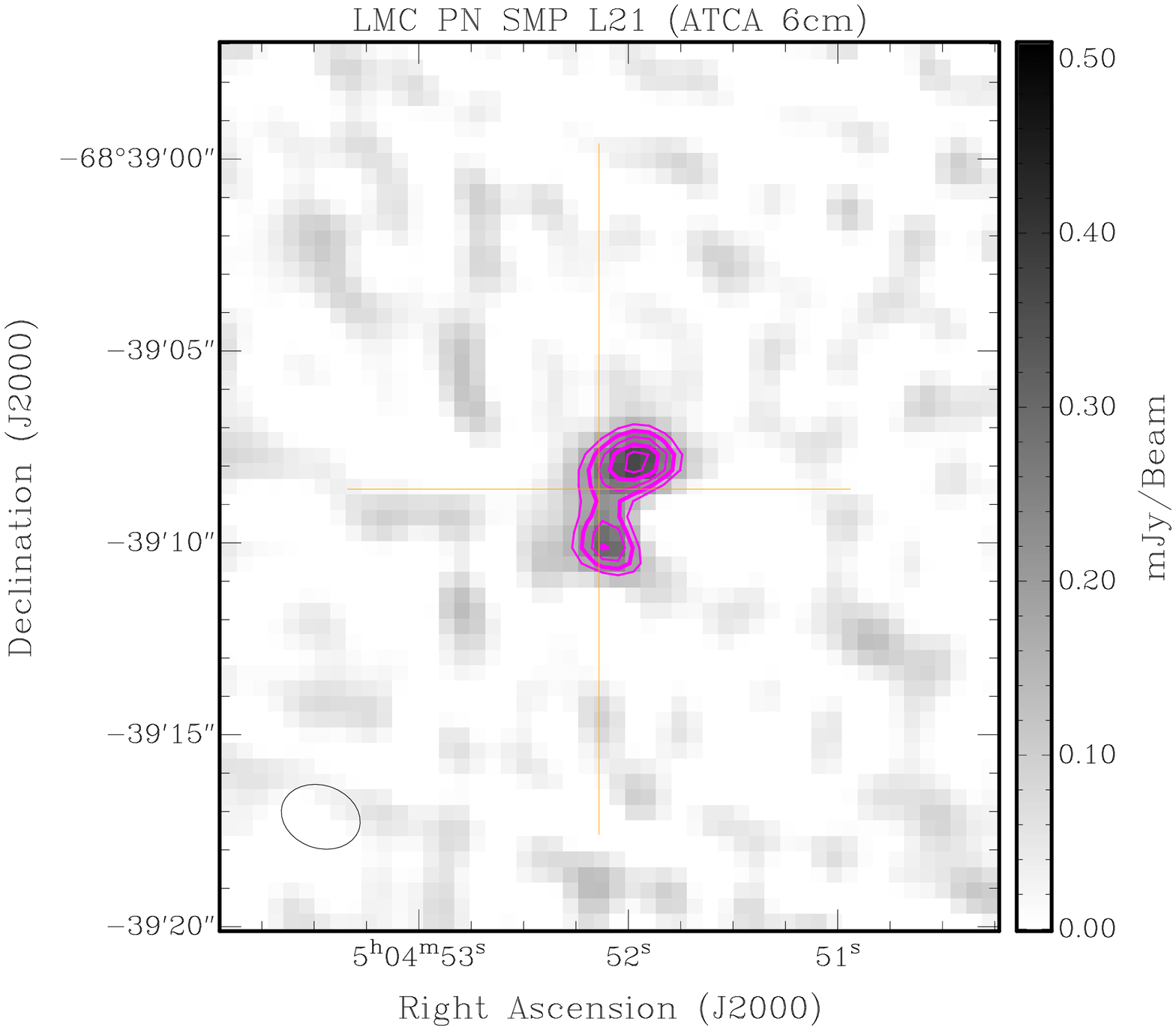}
\includegraphics[angle=-90, trim=0 0 0 0, width=.475\textwidth]{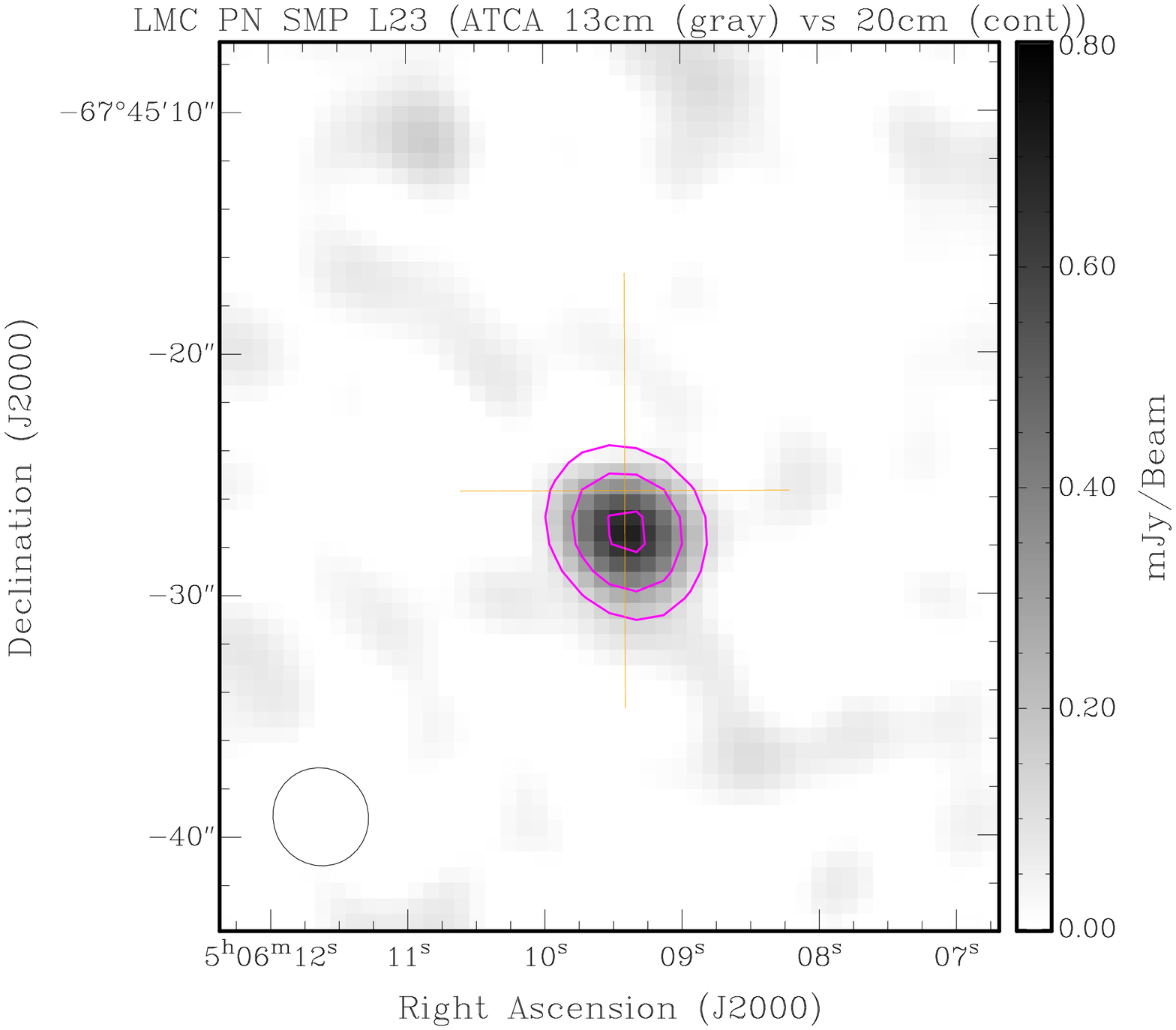}
\includegraphics[angle=-90, trim=0 0 0 0, width=.475\textwidth]{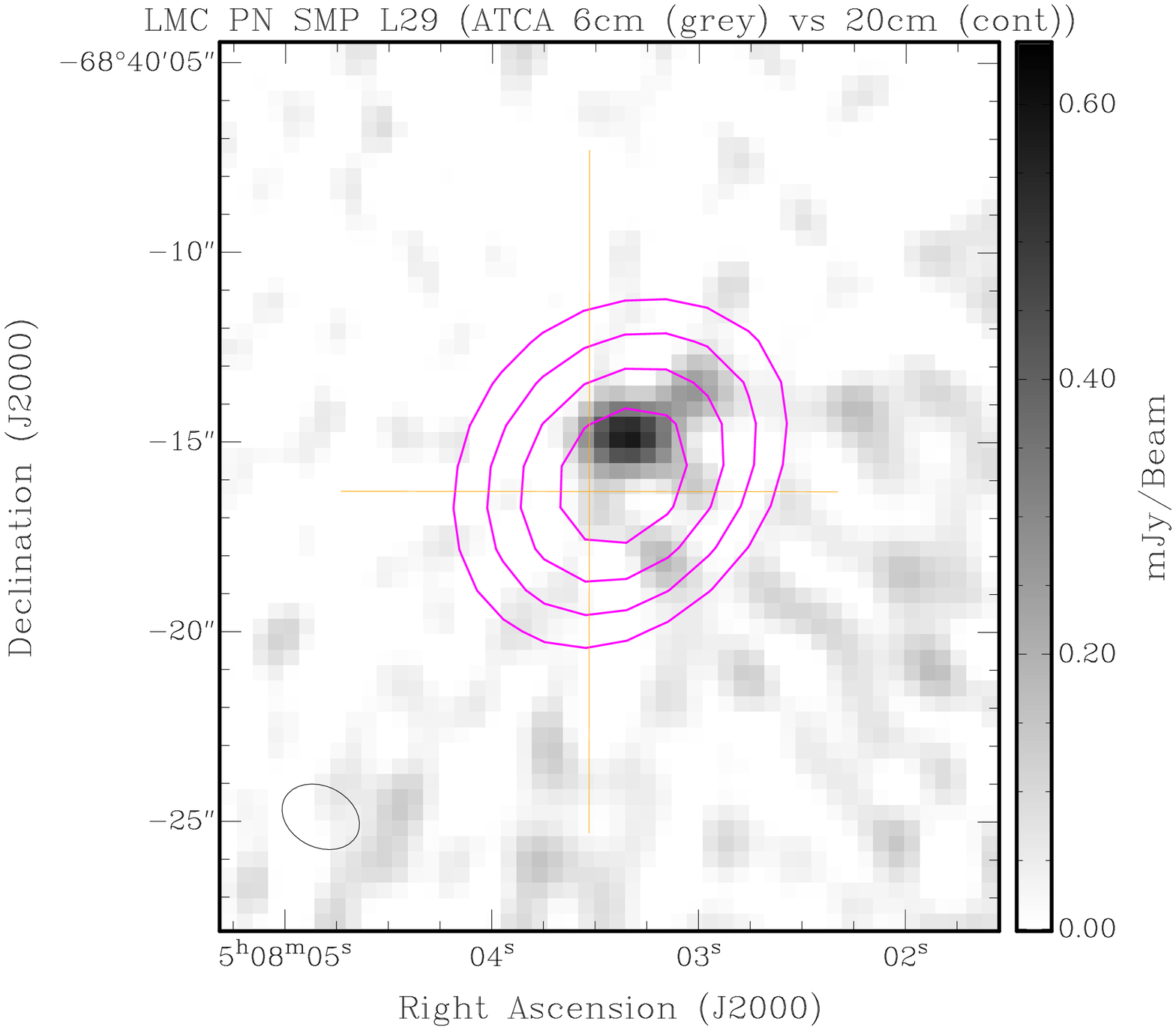}
\includegraphics[angle=-90, trim=0 0 0 0, width=.475\textwidth]{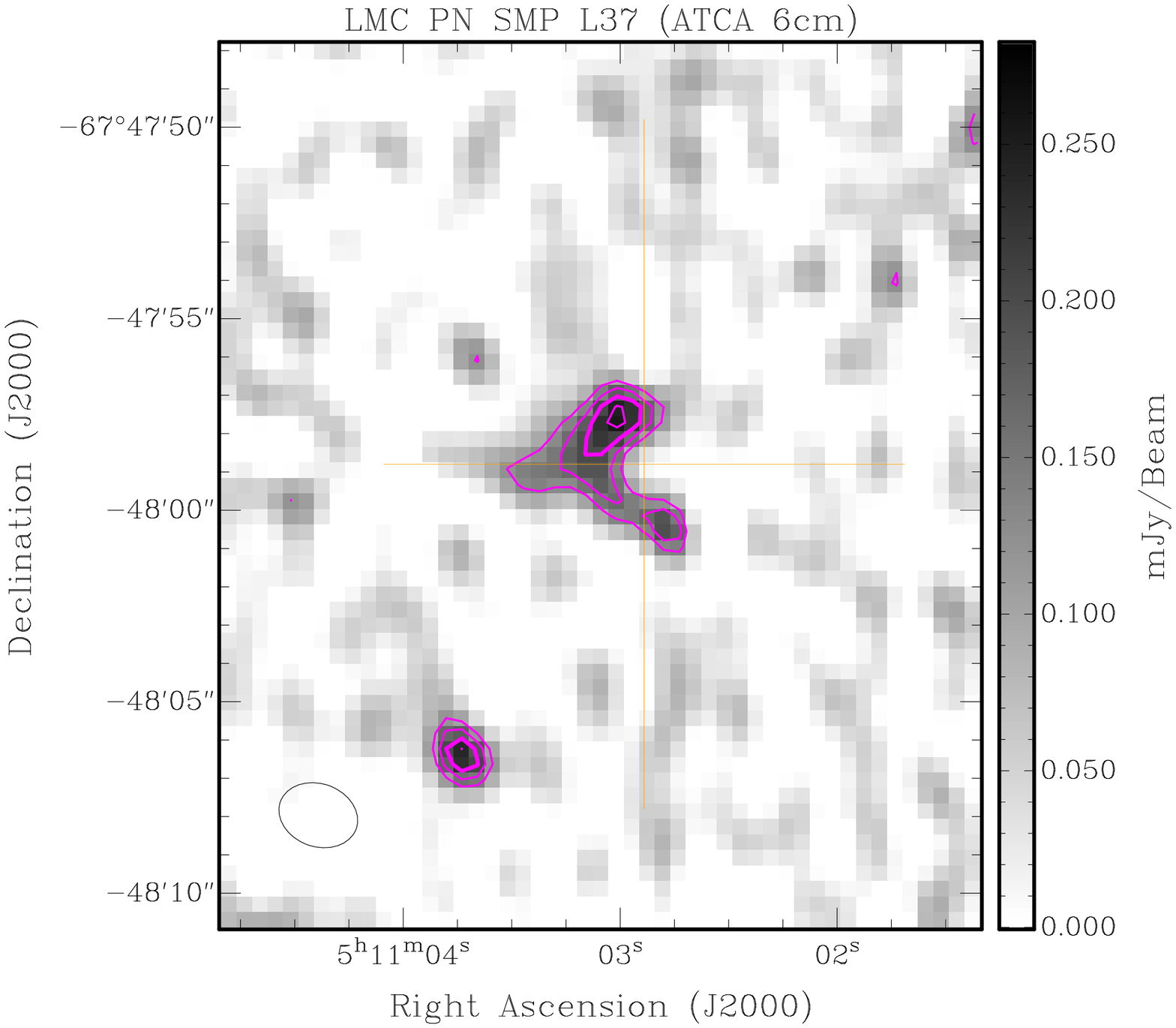}
\caption{New and previous radio detections of LMC PNe. Images are constructed as total intensity radio maps. Contours are integral multiples of the measured RMS noise starting at 3$\sigma$ with spacings of 1$\sigma$. Top row: SMP\,L13 at 6~cm (grey scale) with 20~cm contours and SMP\,L15 at 6~cm (grey scale) with 20~cm contours; Second row: SMP\,L21 at 6~cm and SMP\,L23 at 13~cm (grey scale) with 20~cm contours; Third row: SMP\,L29 at 6~cm (grey scale) with 20~cm contours and SMP\,L37 at 6~cm. The beam size of each image is shown in the bottom left corner. The orange crosses represent the PN positions from the RP catalogue.}
\label{fig:radio_gr1}
\end{figure*}

\begin{figure*}
\centering
\includegraphics[angle=-90, trim=0 0 0 0, width=.475\textwidth]{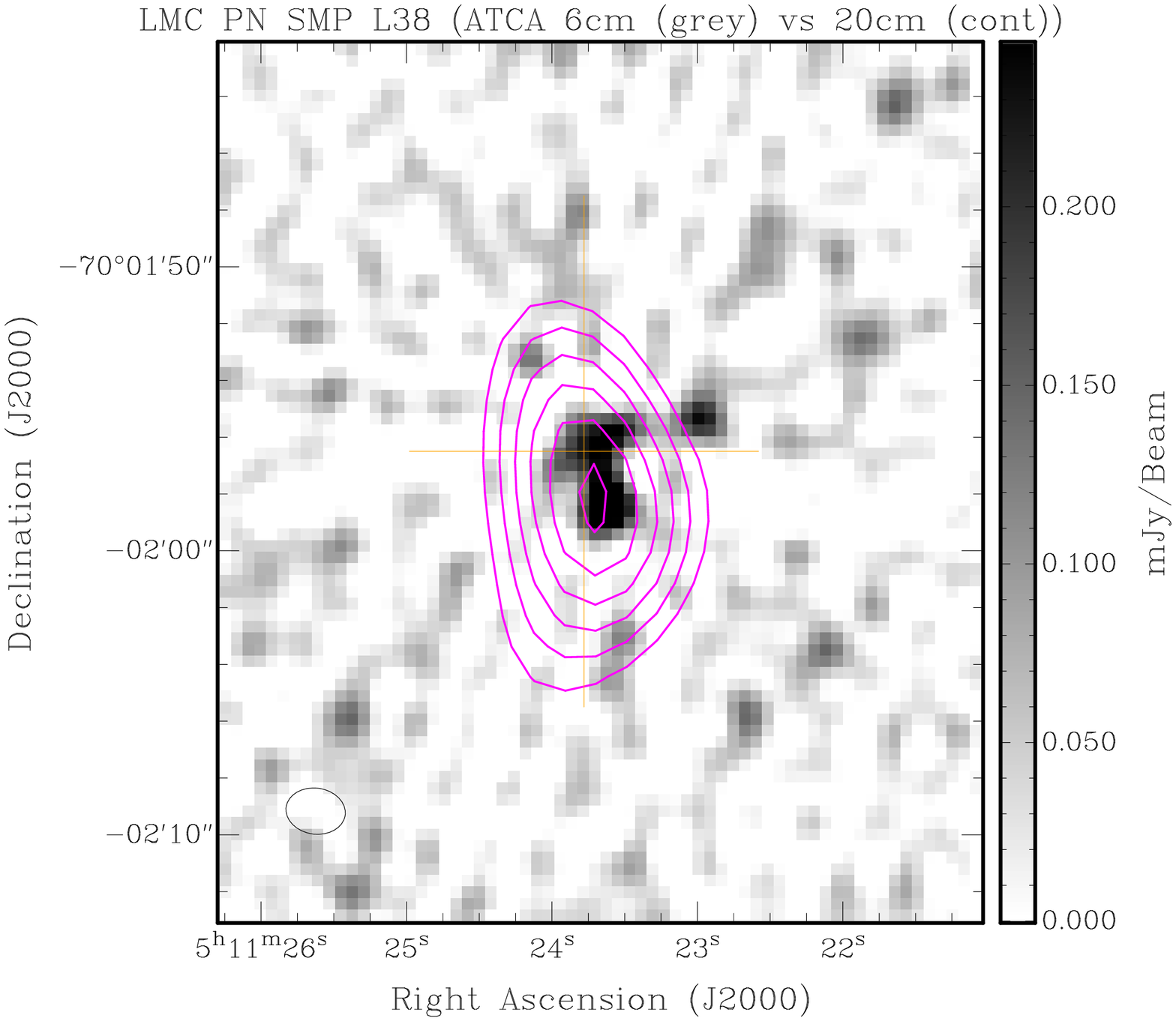}
\includegraphics[angle=-90, trim=0 0 0 0, width=.475\textwidth]{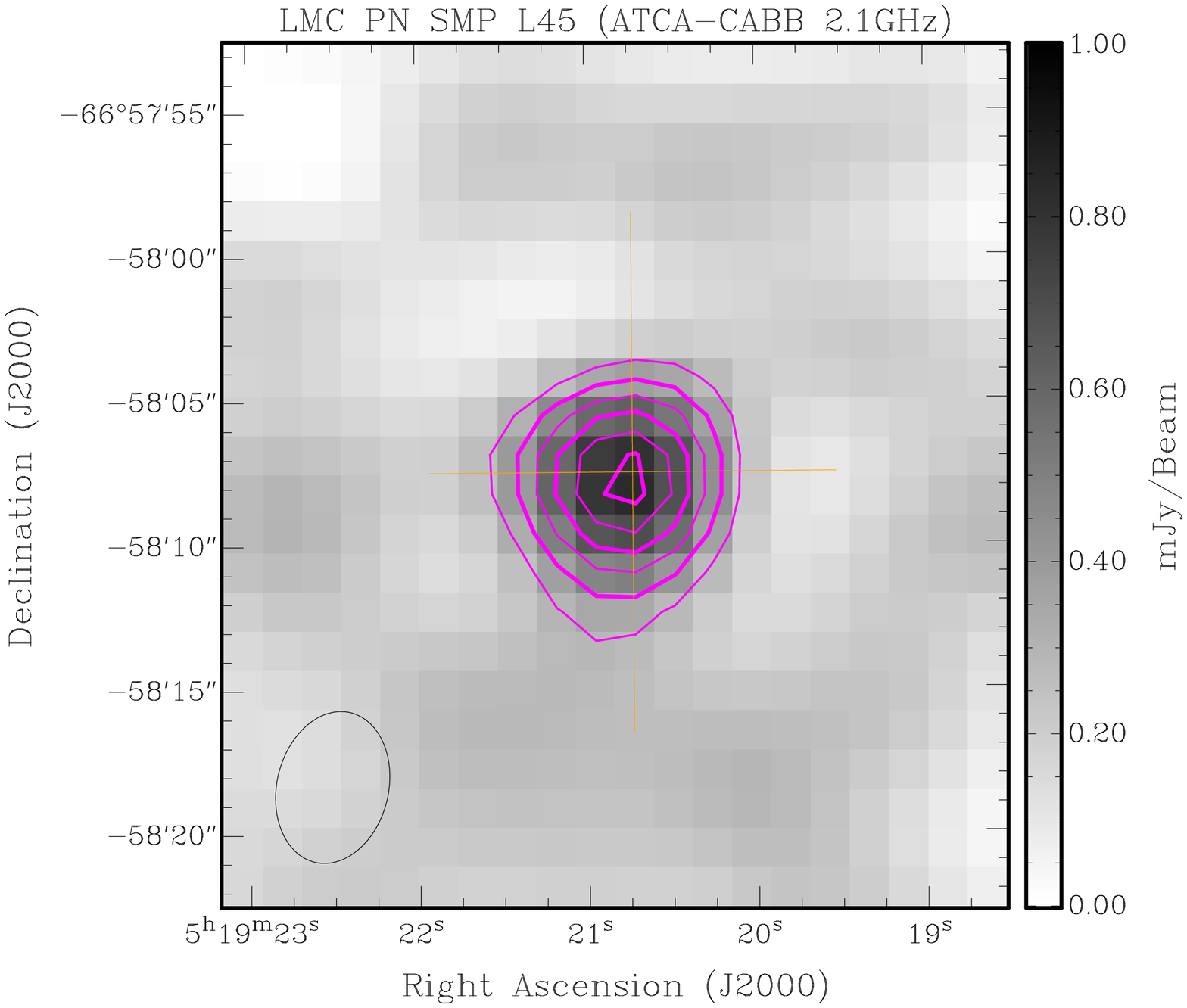}
\includegraphics[angle=-90, trim=0 0 0 0, width=.475\textwidth]{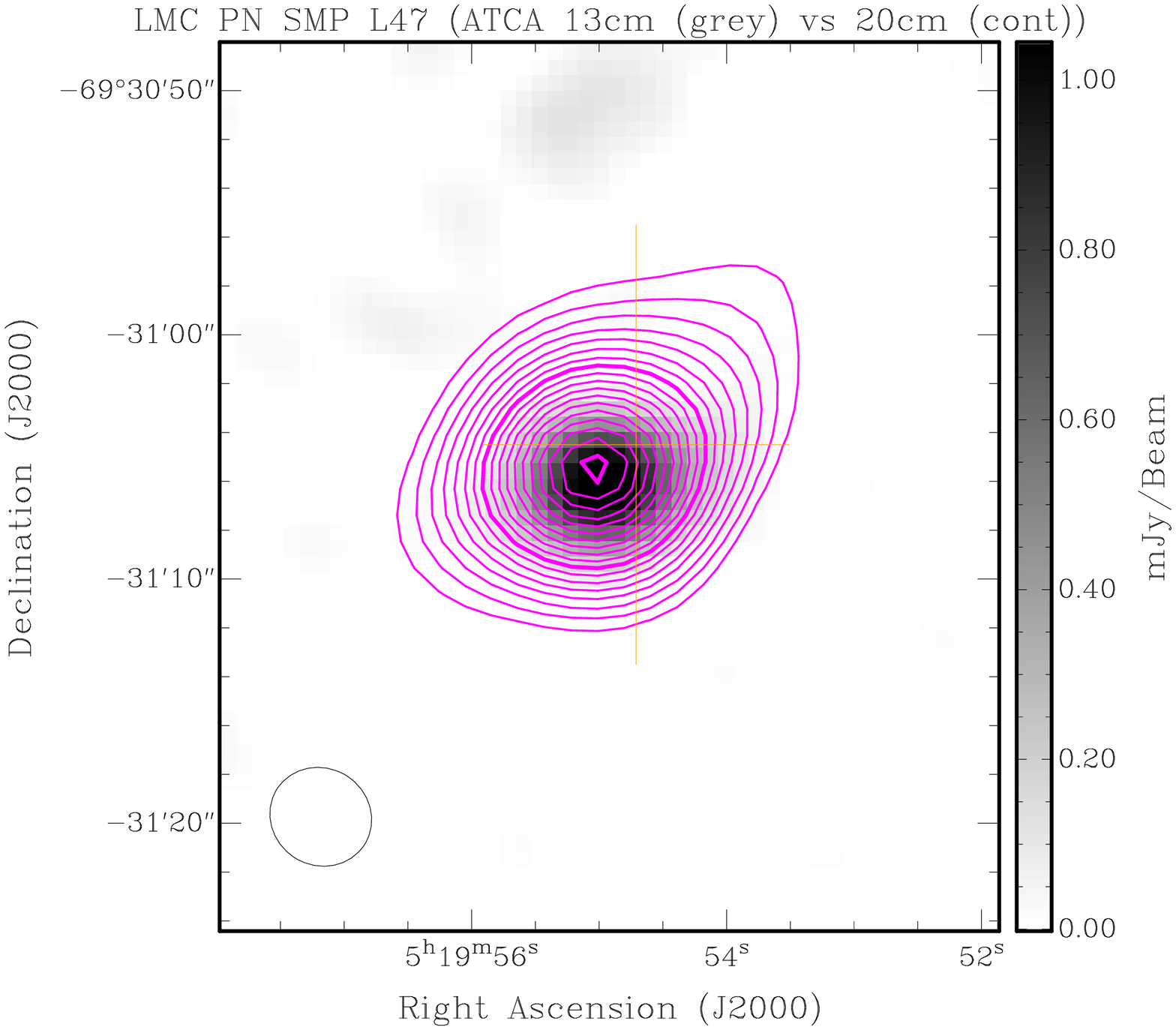}
\includegraphics[angle=-90, trim=0 0 0 0, width=.475\textwidth]{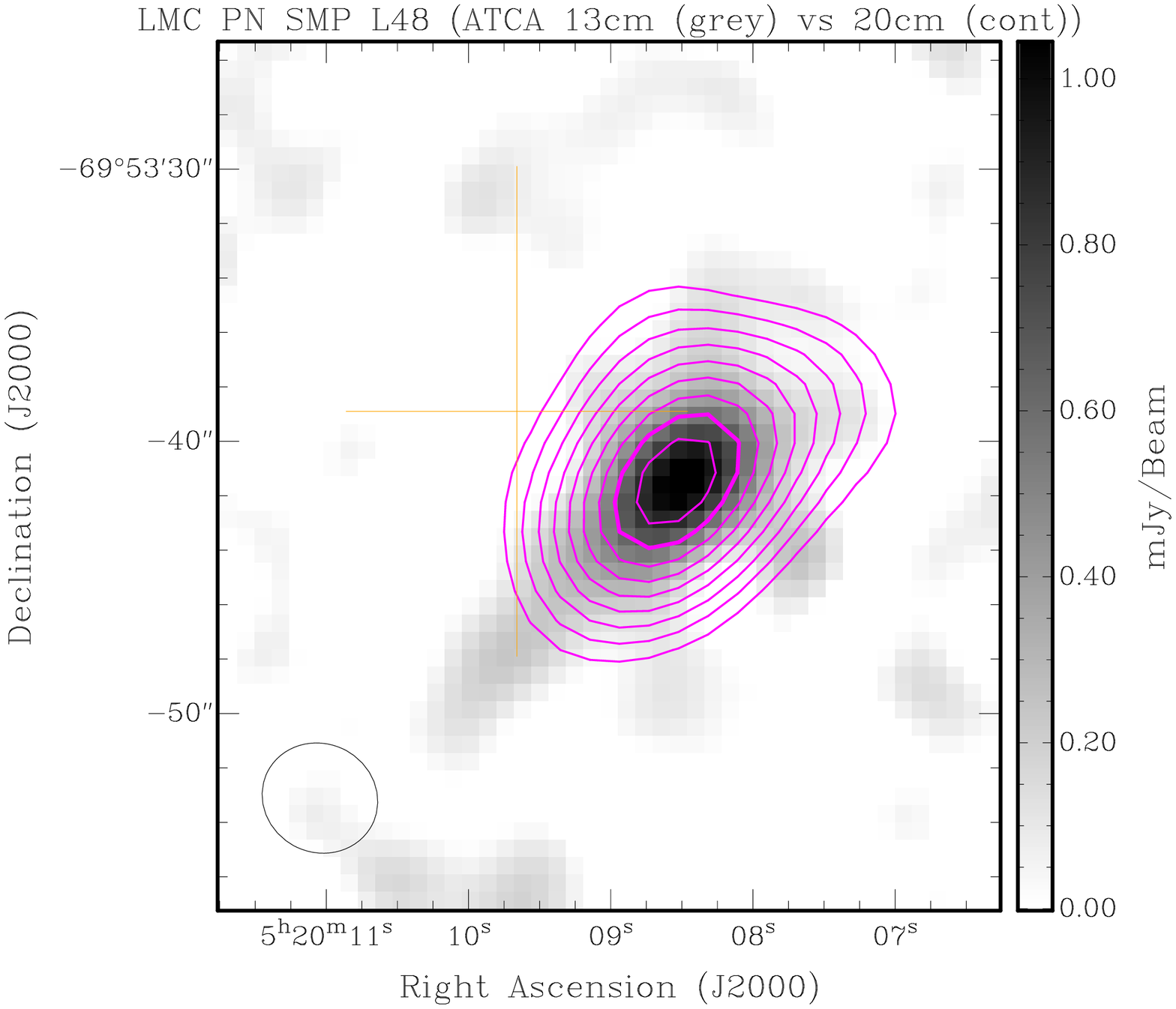}
\includegraphics[angle=-90, trim=0 0 0 0, width=.475\textwidth]{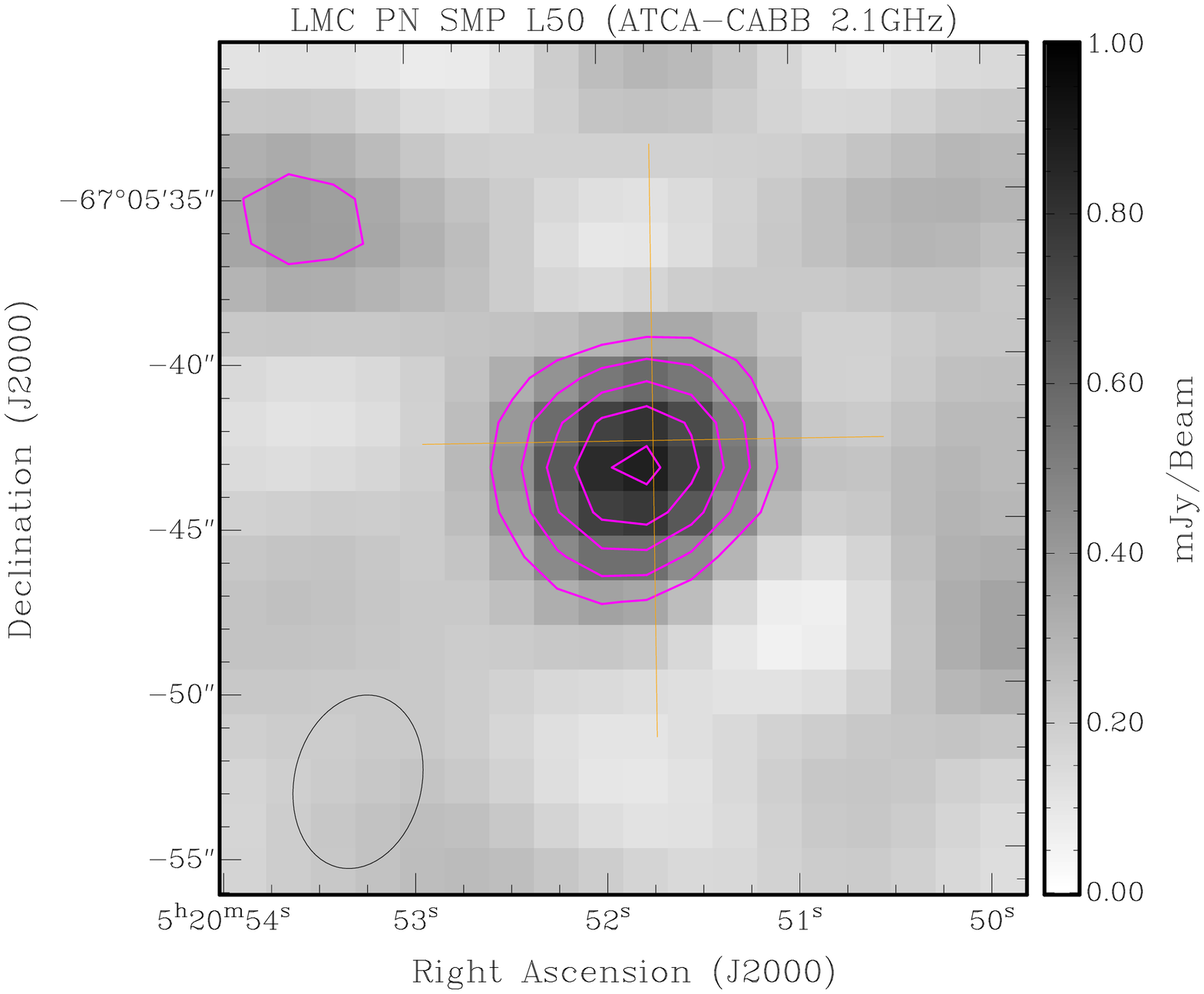}
\includegraphics[angle=-90, trim=0 0 0 0, width=.475\textwidth]{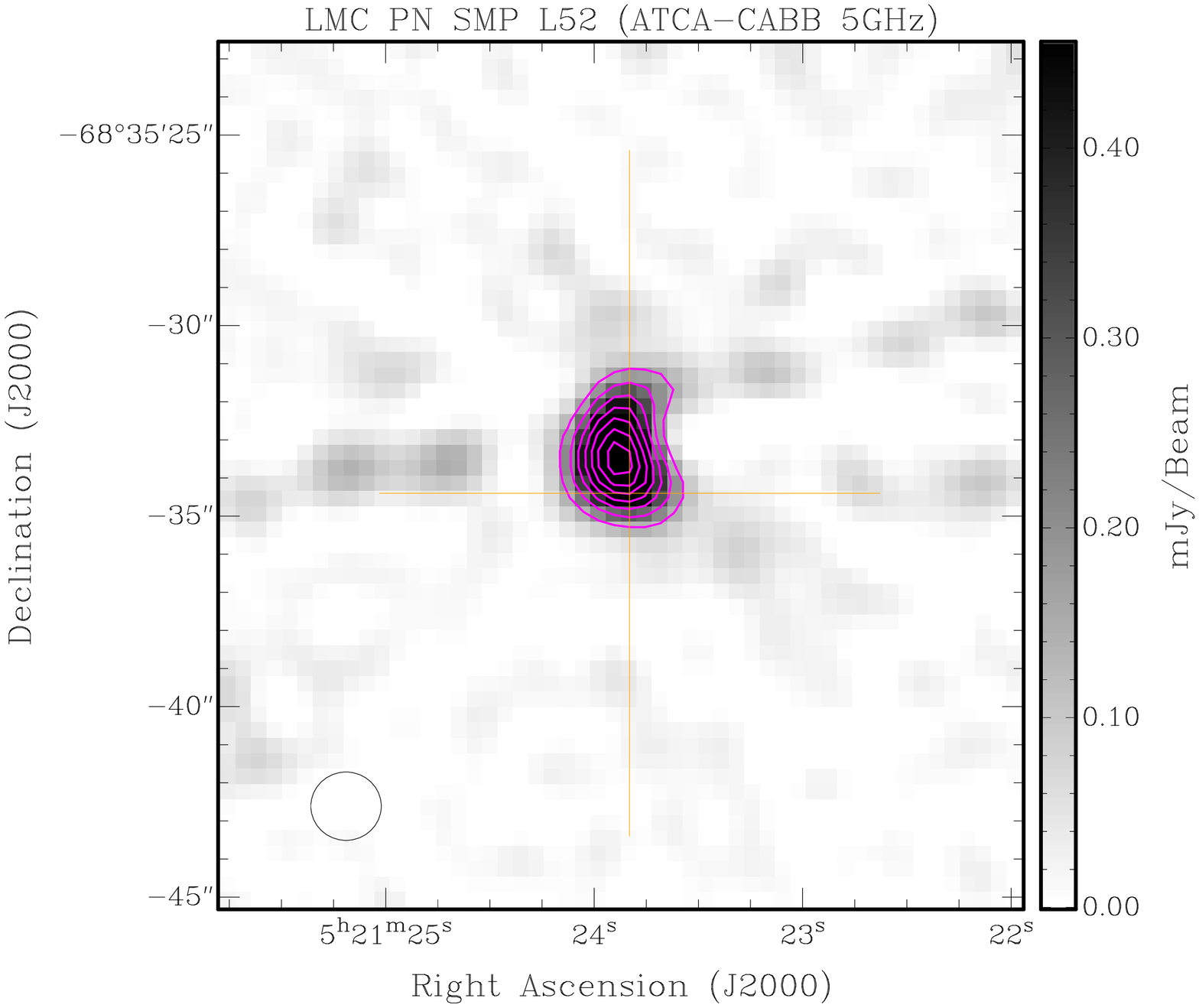}
\caption{New and previous radio detections of LMC PNe. Images are constructed as total intensity radio maps. Contours are integral multiples of the measured RMS noise starting at 3$\sigma$ with spacings of 1$\sigma$. Top row: SMP\,L38 at 6~cm (grey scale) with 20~cm contours and SMP\,L45 at 2.1~GHz; Second row: SMP\,L47 and SMP\,48 at 13~cm (grey scale) with 20~cm contour; Third row: SMP\,L50 at 2.1~GHz and SMP\,L52 at 5~GHz. The beam size of each image is shown in the bottom left corner. The orange crosses represent the PN positions from the RP catalogue.}
\label{fig:radio_gr2}
\end{figure*}

\begin{figure*}
\centering
\includegraphics[angle=-90, trim=0 0 0 0, width=.475\textwidth]{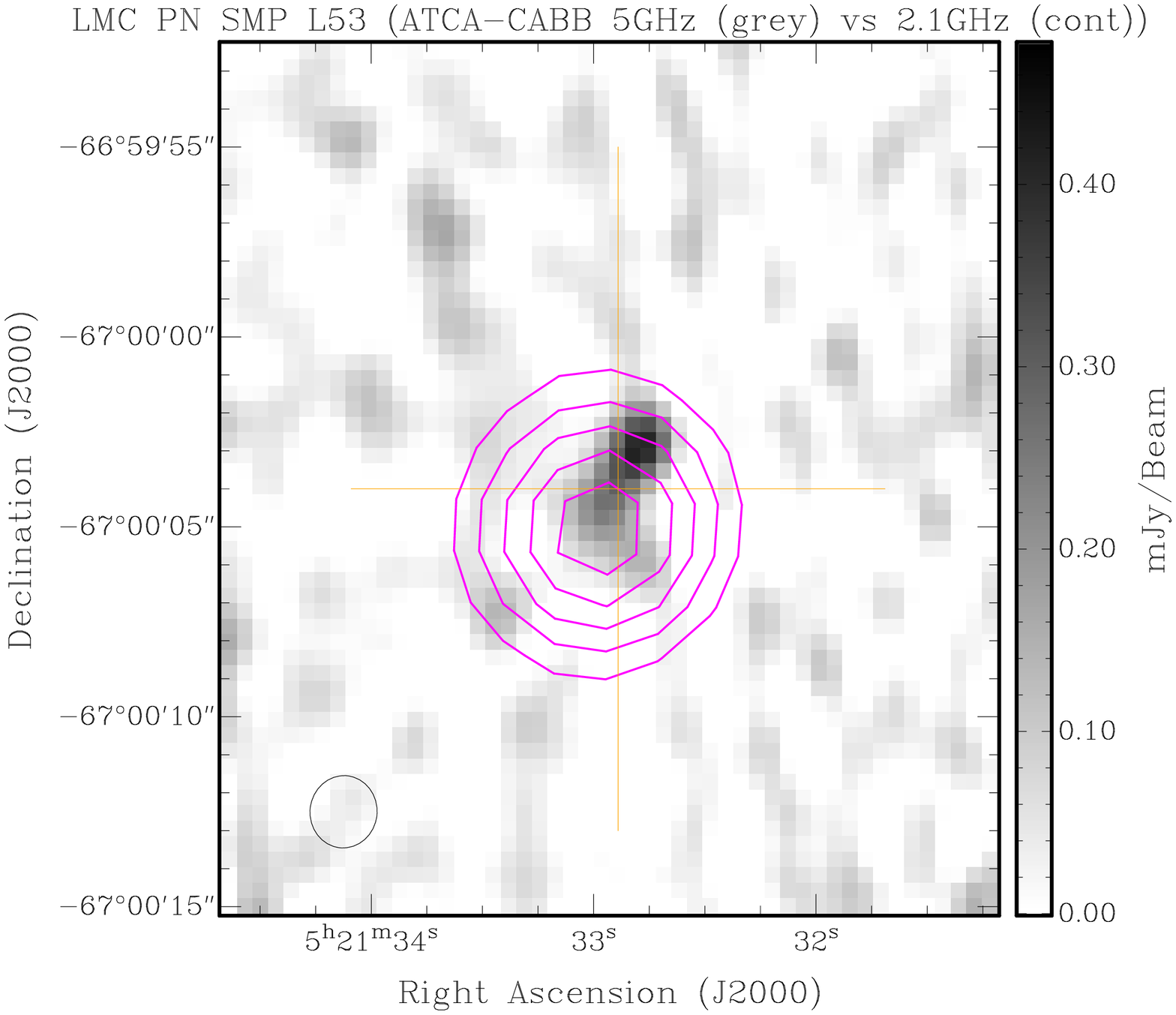}
\includegraphics[angle=-90, trim=0 0 0 0, width=.475\textwidth]{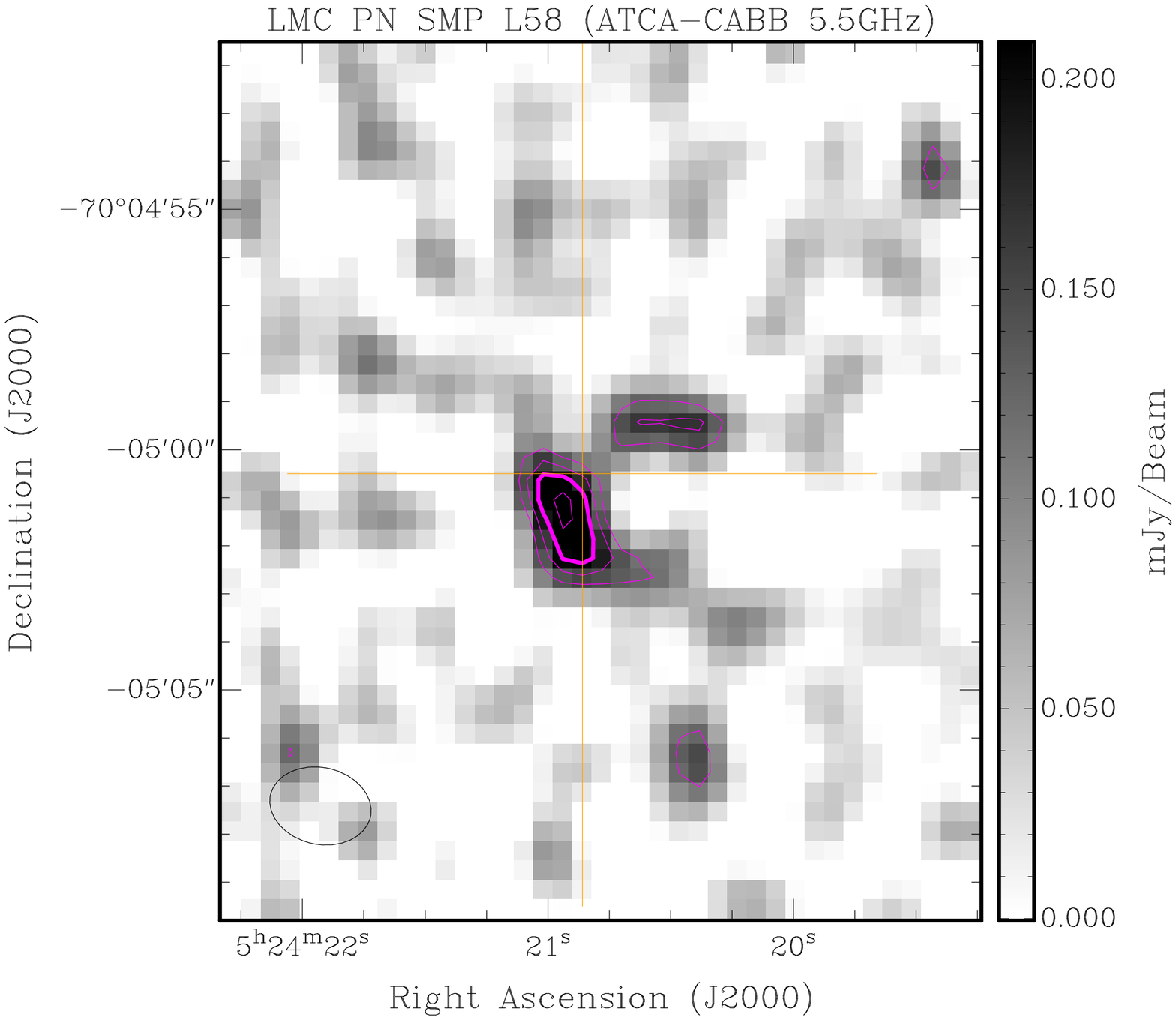}
\includegraphics[angle=-90, trim=0 0 0 0, width=.475\textwidth]{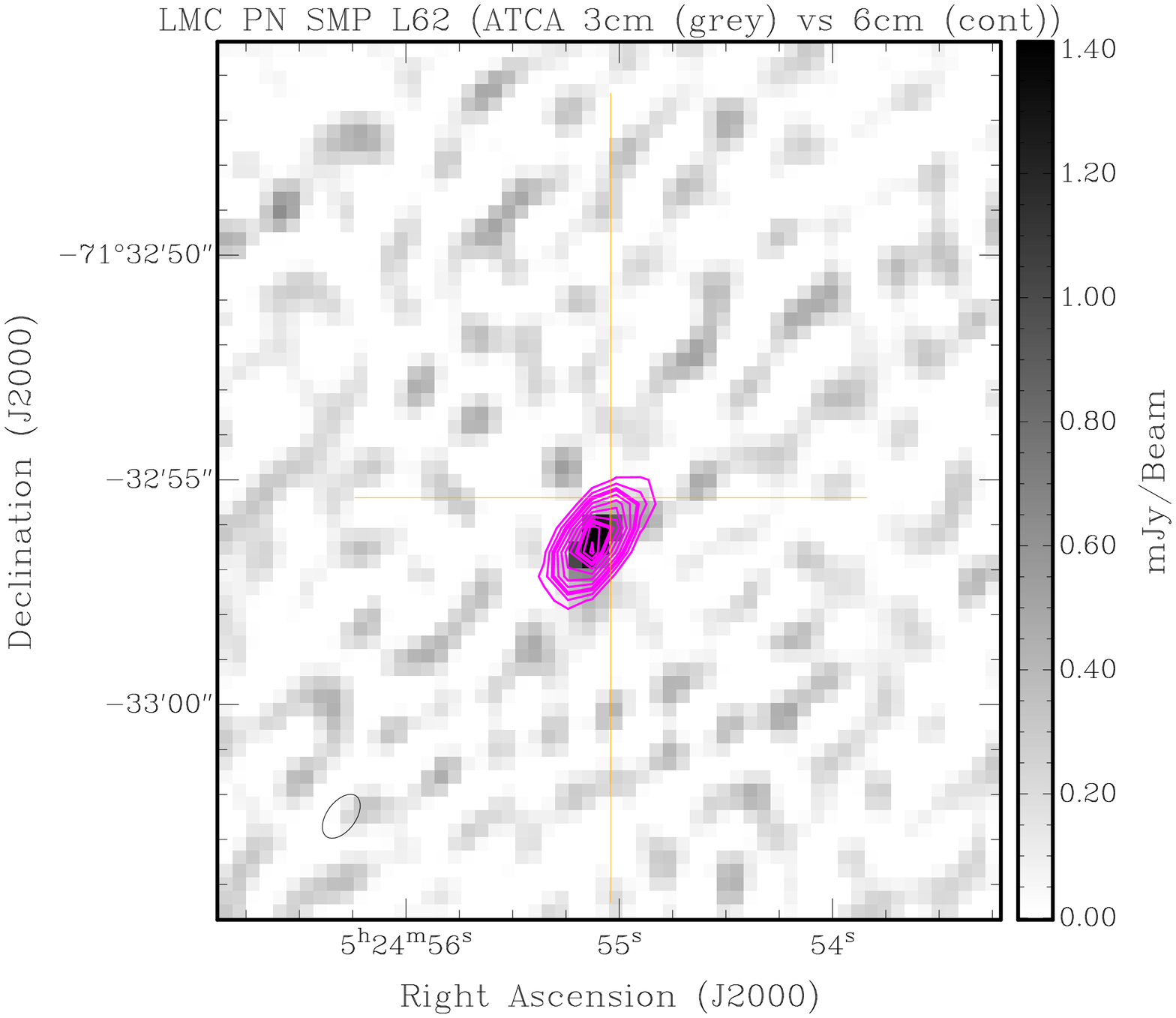}
\includegraphics[angle=-90, trim=0 0 0 0, width=.475\textwidth]{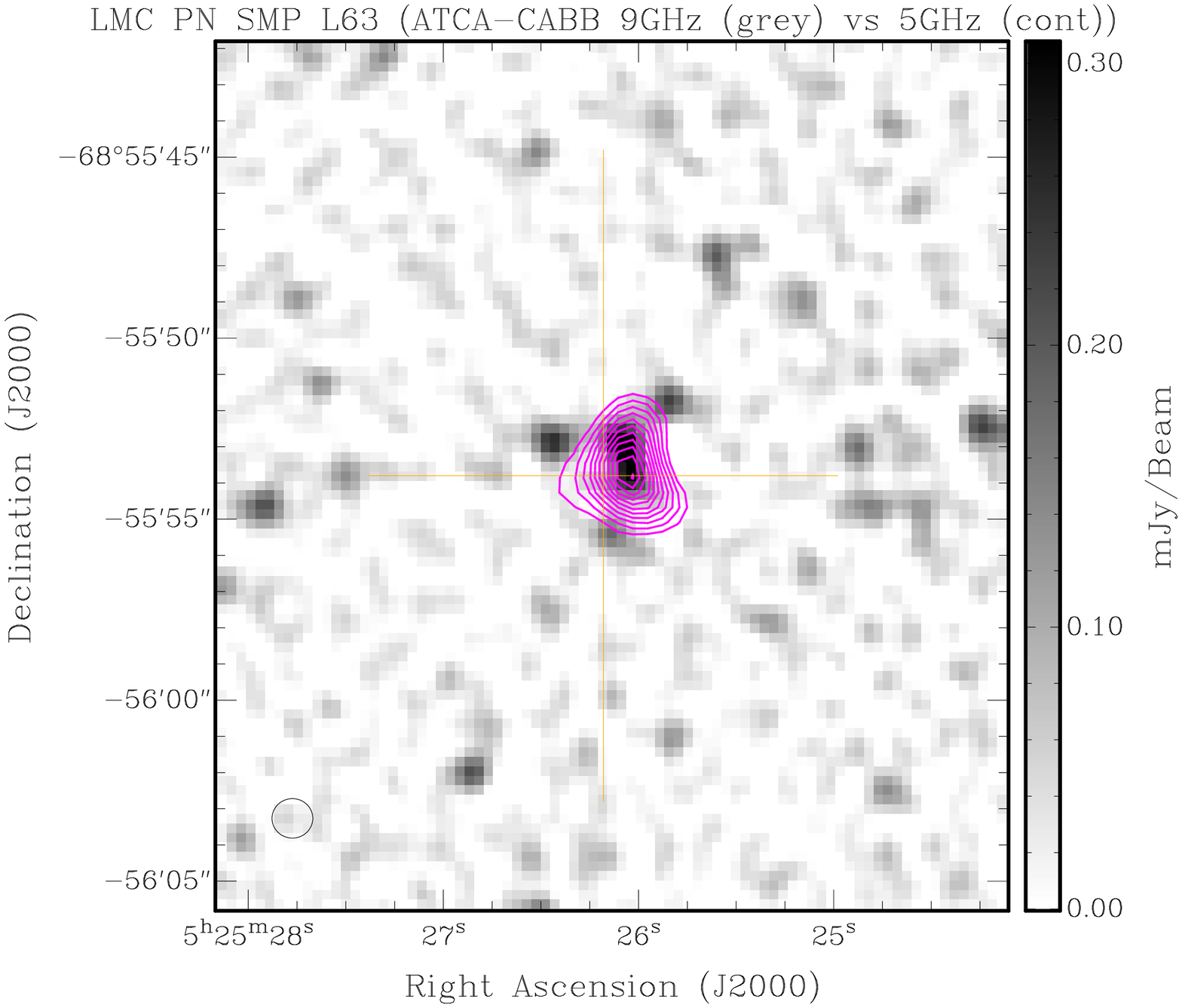}
\includegraphics[angle=-90, trim=0 0 0 0, width=.475\textwidth]{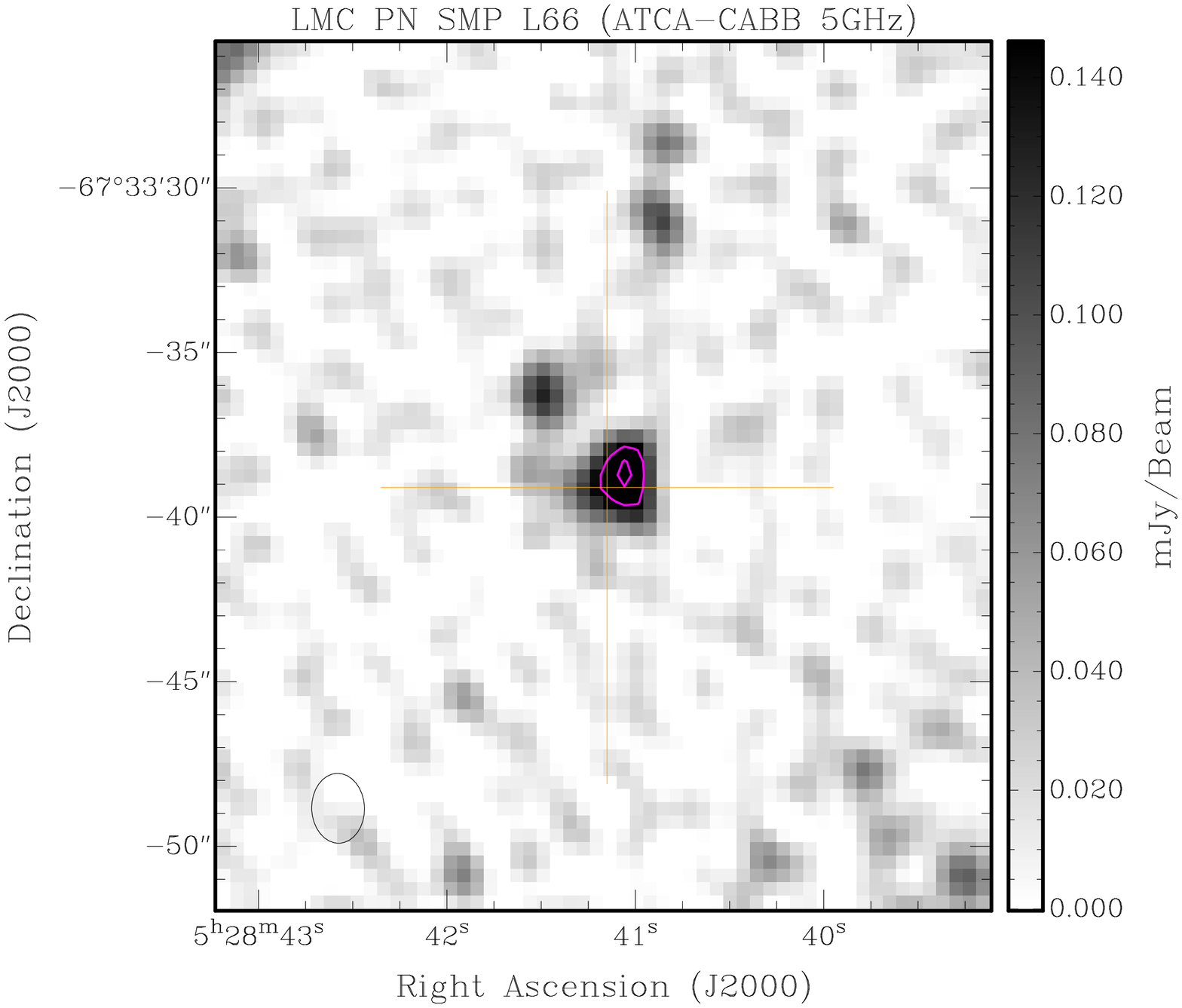}
\includegraphics[angle=-90, trim=0 0 0 0, width=.475\textwidth]{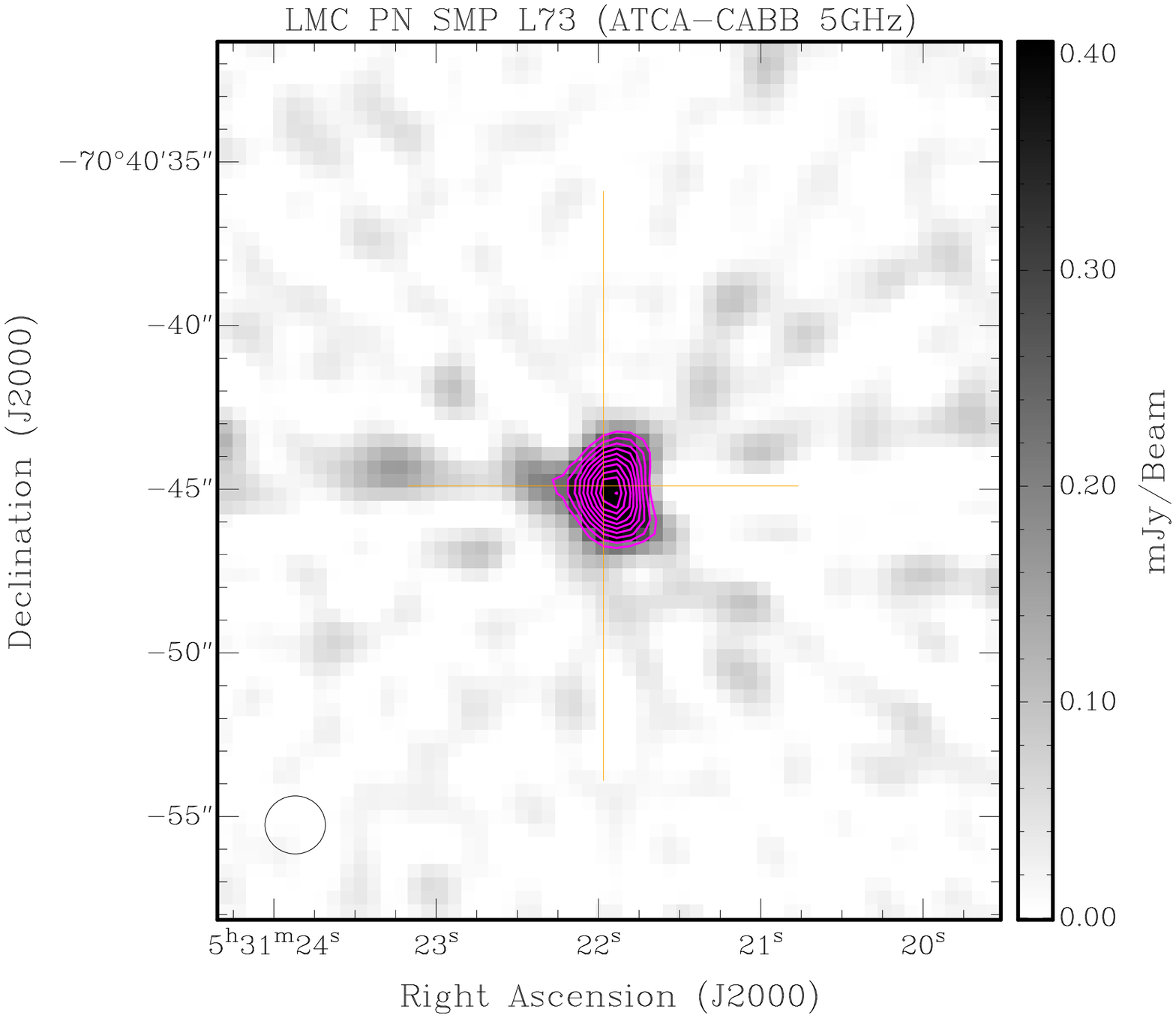}
\caption{New and previous radio detections of LMC PNe. Images are constructed as total intensity radio maps. Contours are integral multiples of the measured RMS noise starting at 3$\sigma$ with spacings of 1$\sigma$. Top row: SMP\,L53 at 5~GHz (grey scale) with 2.1~GHz contours and SMP\,L58 at 5.5~GHz; Second row: SMP\,L62 at 3~cm (grey scale) with 6~cm contours and SMP\,L63 at 9~GHz (grey scale) with 5~GHz contours; Third row: SMP\,L66 at 5~GHz and SMP\,L73 at 5~GHz. The beam size of each image is shown in the bottom left corner. The orange crosses represent the PN positions from the RP catalogue.}
\label{fig:radio_gr3}
\end{figure*}

\begin{figure*}
\centering
\includegraphics[angle=-90, trim=0 0 0 0, width=.475\textwidth]{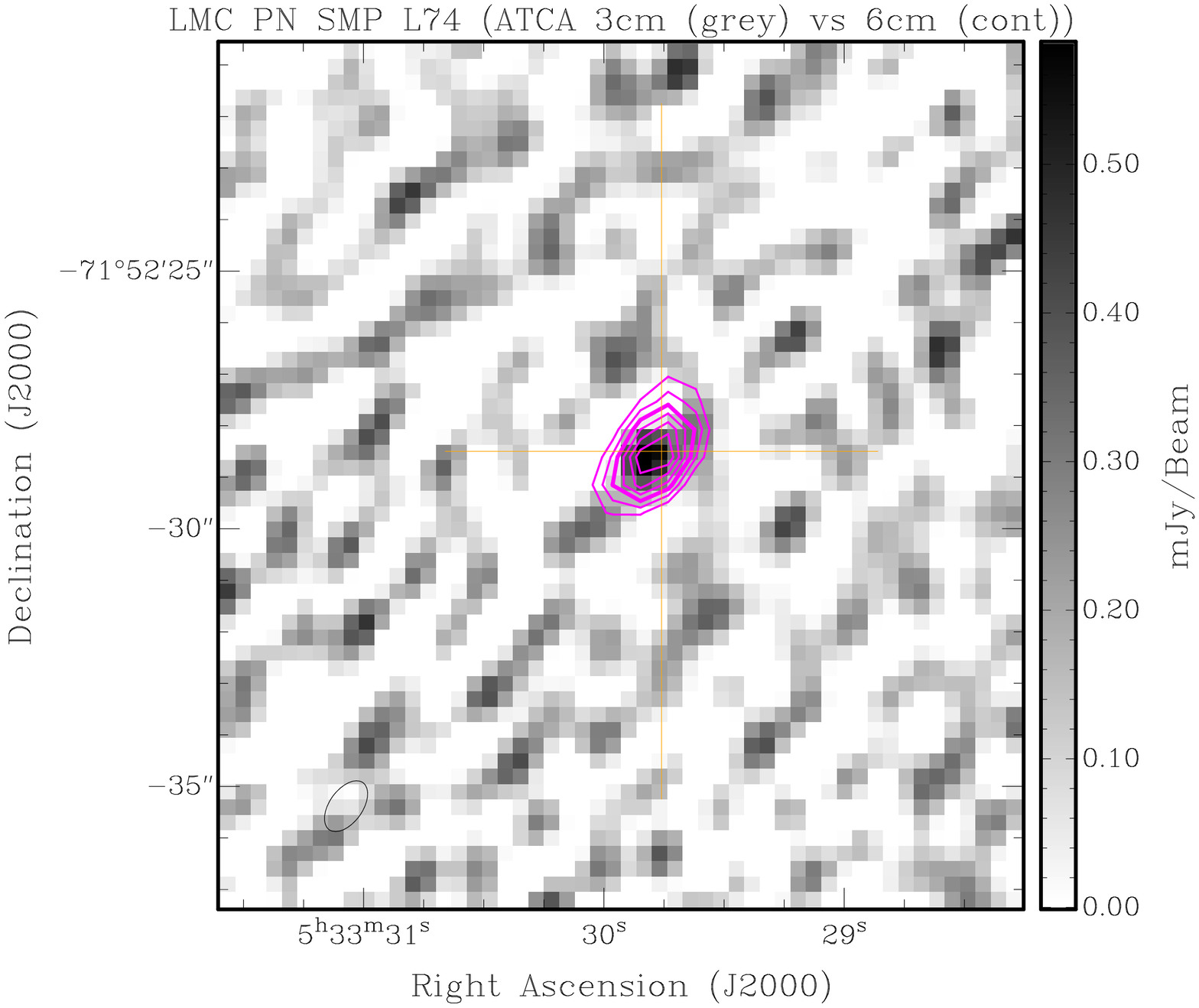}
\includegraphics[angle=-90, trim=0 0 0 0, width=.475\textwidth]{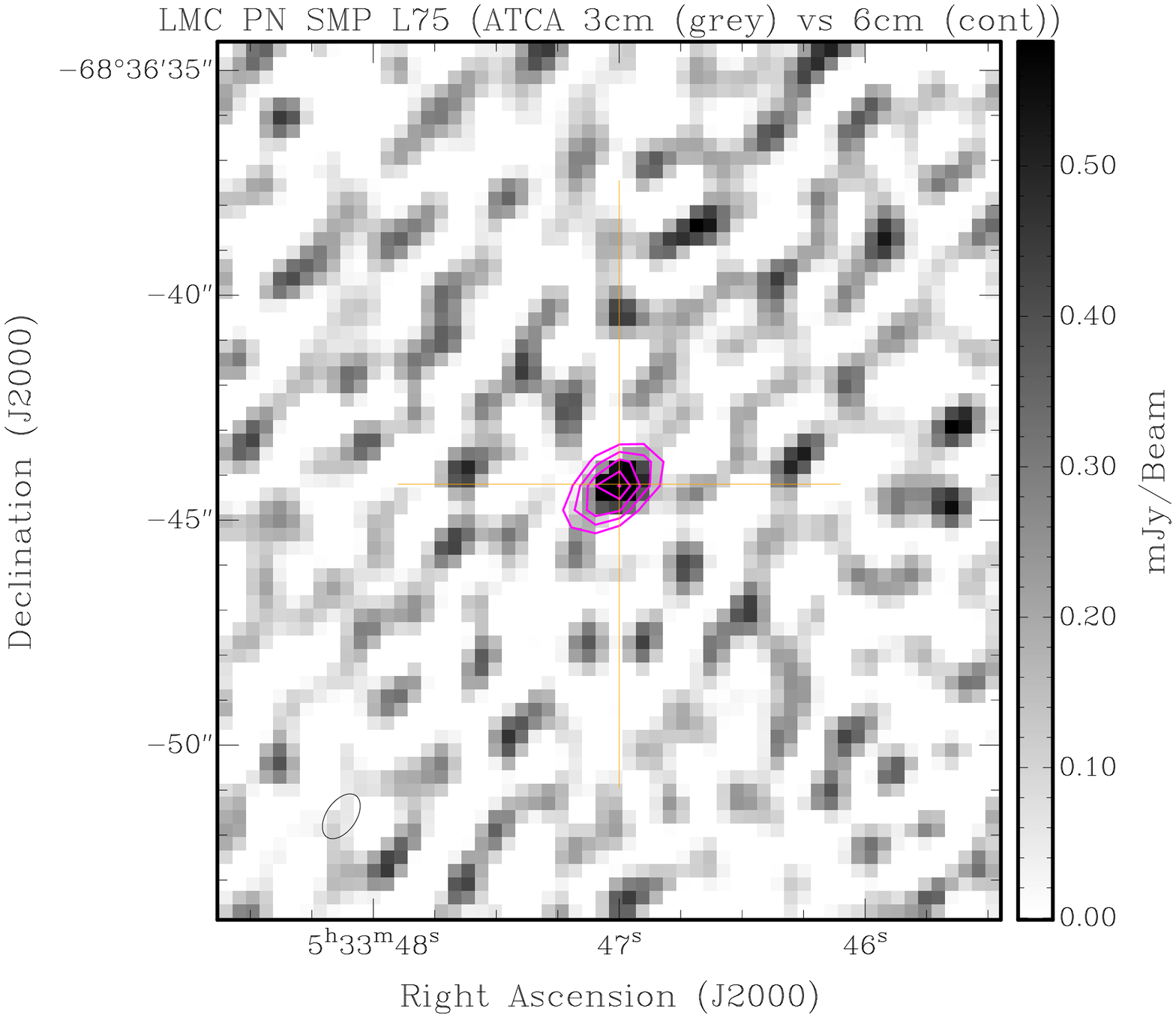}
\includegraphics[angle=-90, trim=0 0 0 0, width=.475\textwidth]{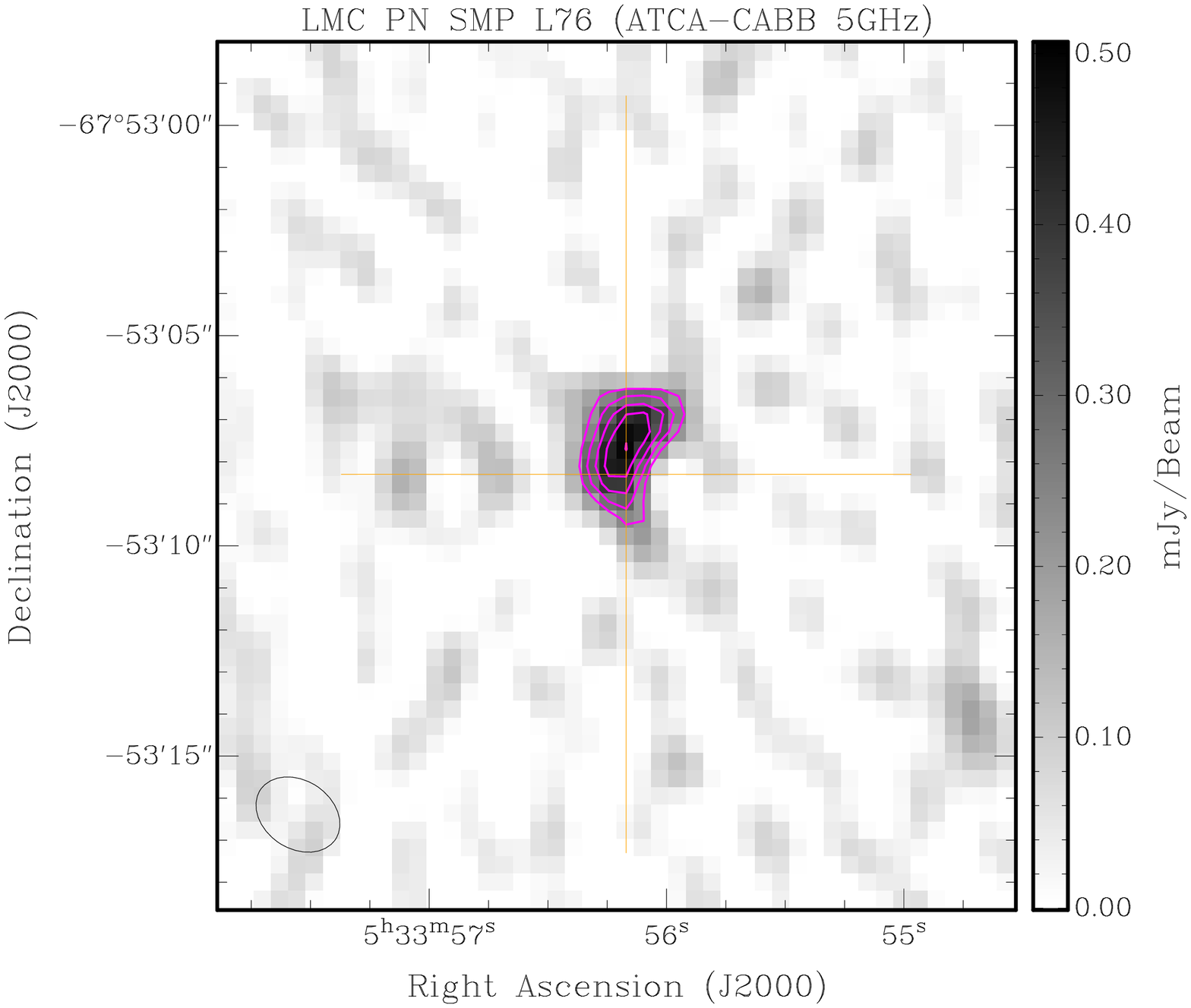}
\includegraphics[angle=-90, trim=0 0 0 0, width=.475\textwidth]{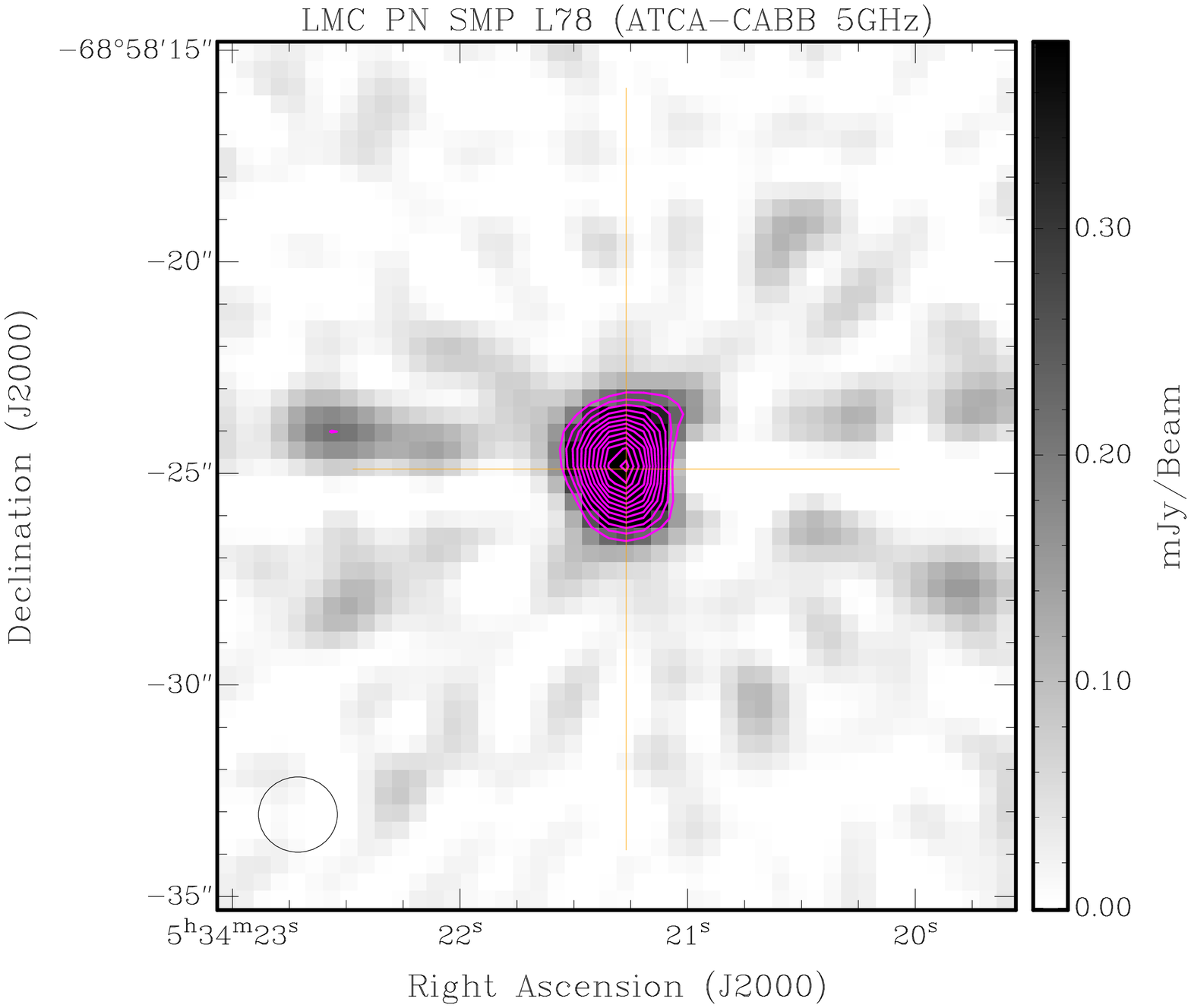}
\includegraphics[angle=-90, trim=0 0 0 0, width=.475\textwidth]{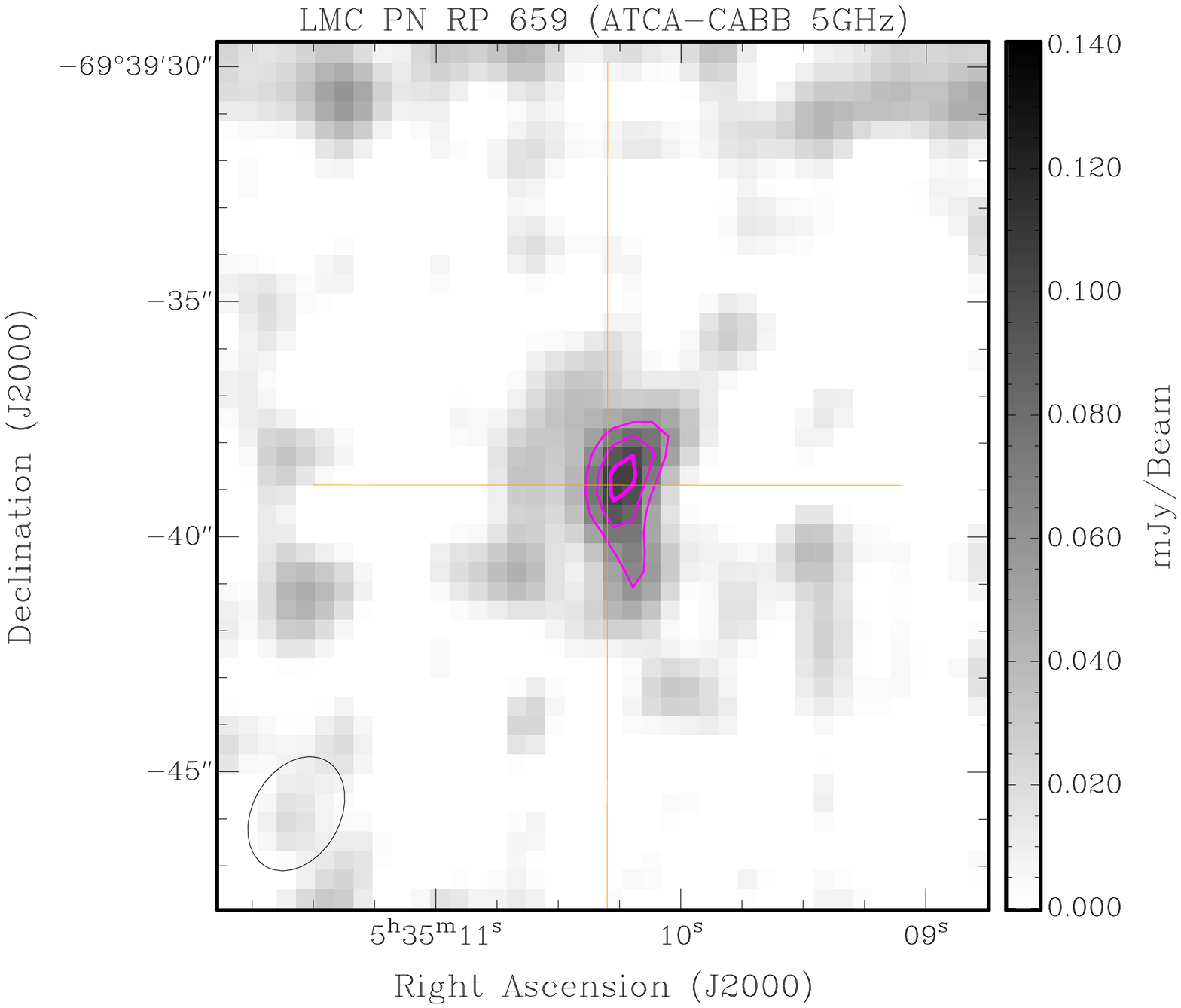}
\includegraphics[angle=-90, trim=0 0 0 0, width=.475\textwidth]{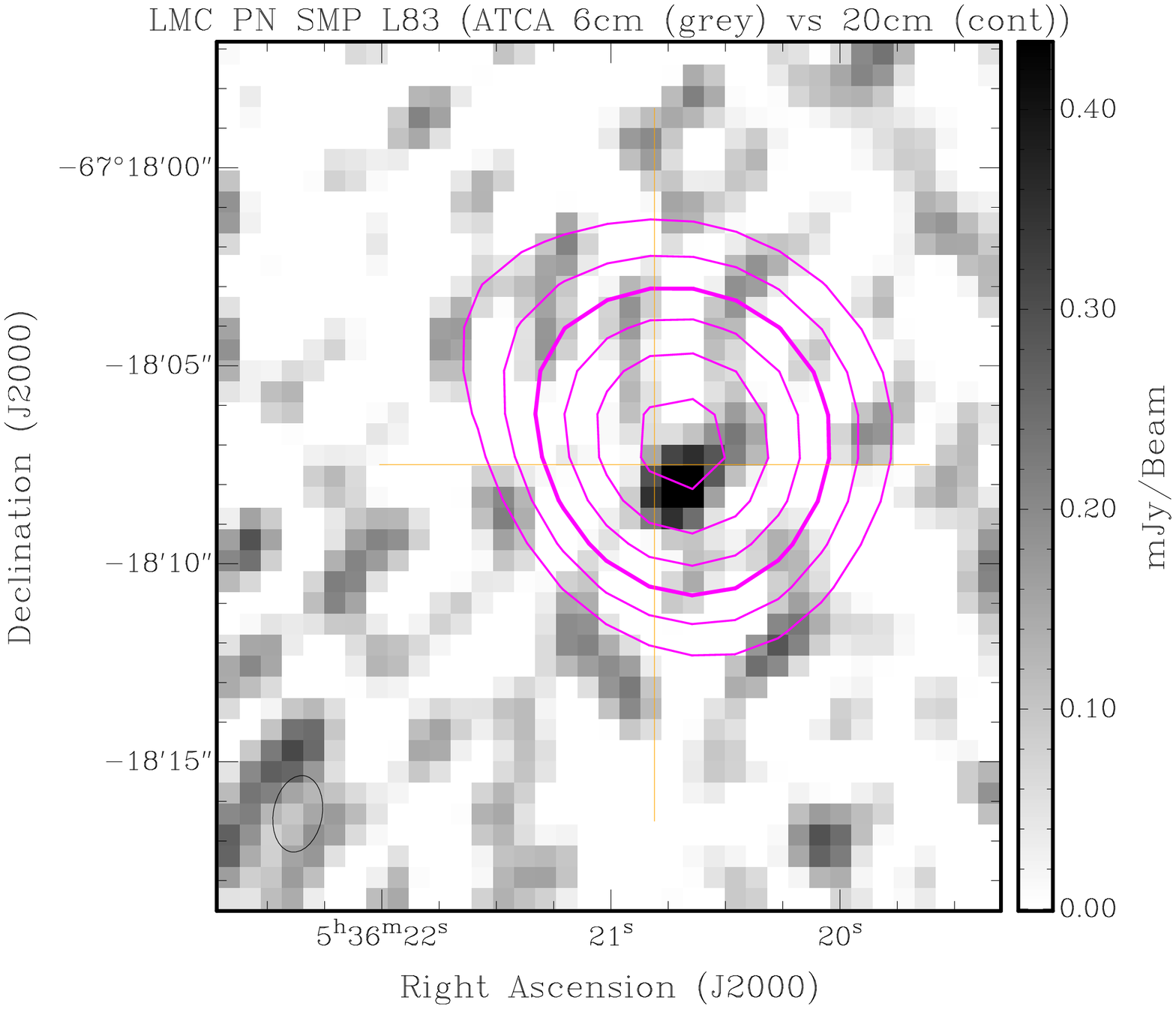}
\caption{New and previous radio detections of LMC PNe. Images are constructed as total intensity radio maps. Contours are integral multiples of the measured RMS noise starting at 3$\sigma$ with spacings of 1$\sigma$. Top row: SMP\,L74 at 3~cm (grey scale) with 6~cm contours and SMP\,L75 at 3~cm (grey scale) with 6~cm contours; Second row: SMP\,L76 at 5~GHz and SMP\,L78 at 5~GHz; Third row: RP\,659 at 5~GHz and SMP\,L83 at 6~cm (grey scale) with 20~cm contours. The beam size of each image is shown in the bottom left corner. The orange crosses represent the PN positions from the RP catalogue.}
\label{fig:radio_gr4}
\end{figure*}

\begin{figure*}
\centering
\includegraphics[angle=-90, trim=0 0 0 0, width=.475\textwidth]{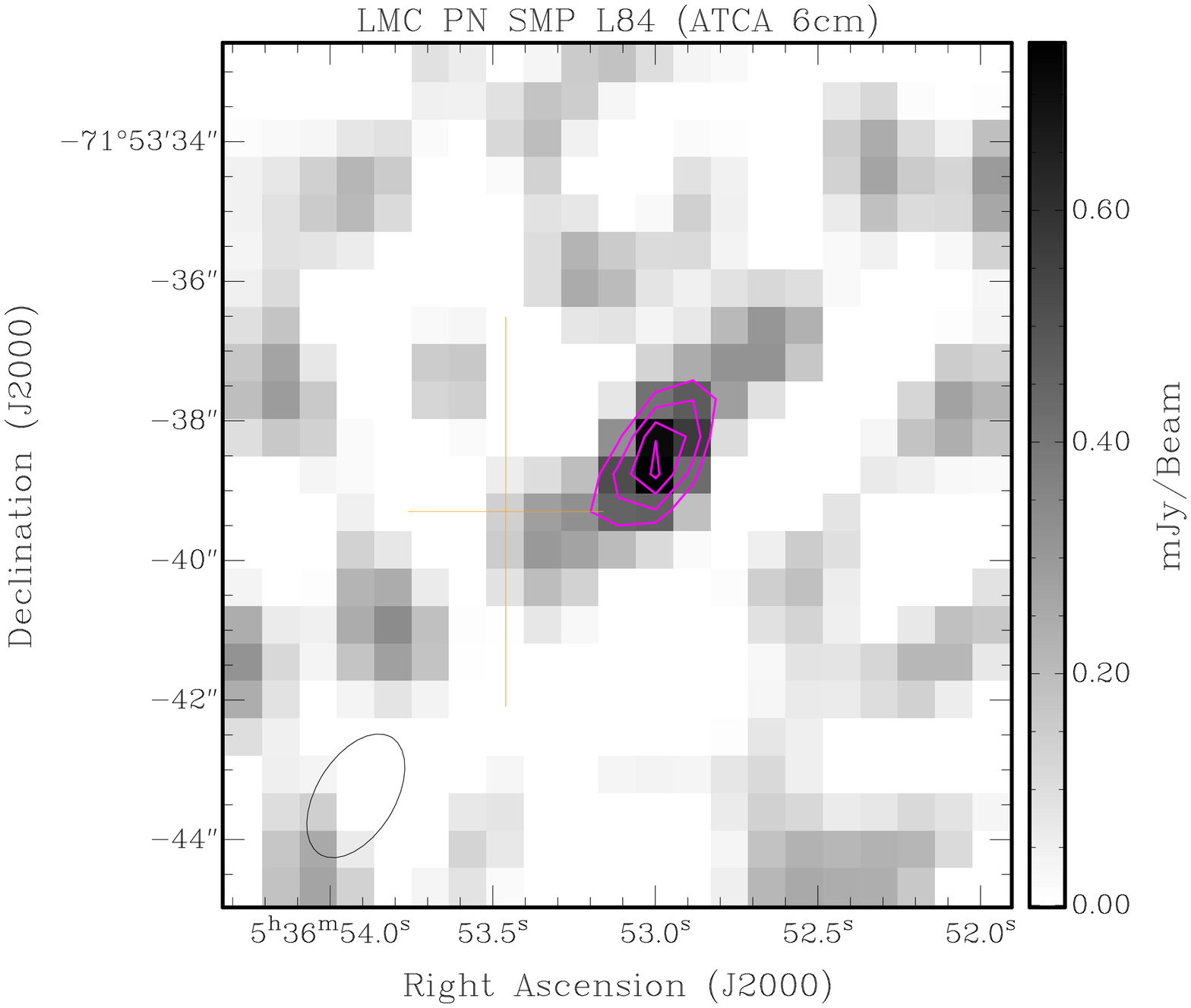}
\includegraphics[angle=-90, trim=0 0 0 0, width=.475\textwidth]{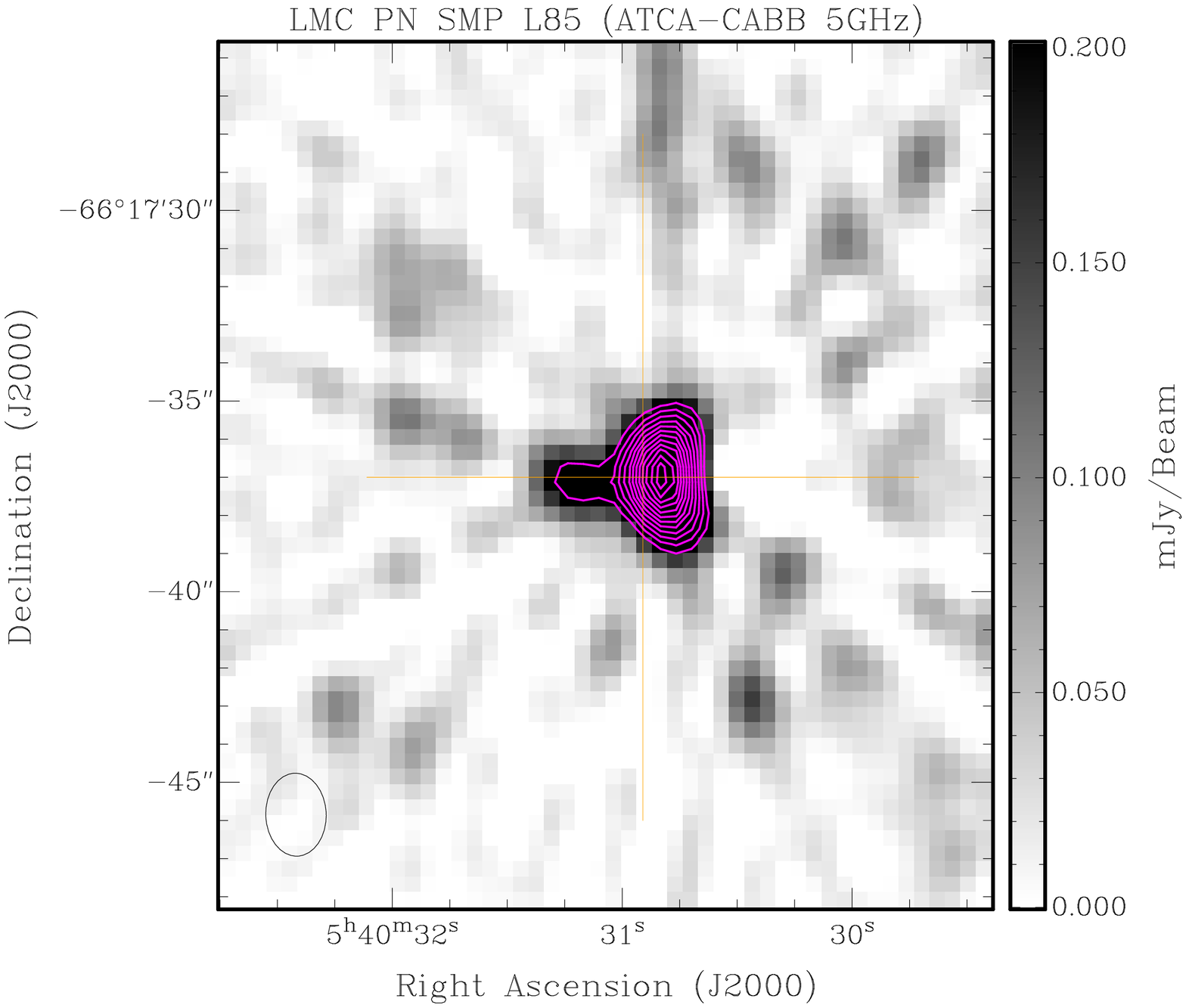}
\includegraphics[angle=-90, trim=0 0 0 0, width=.475\textwidth]{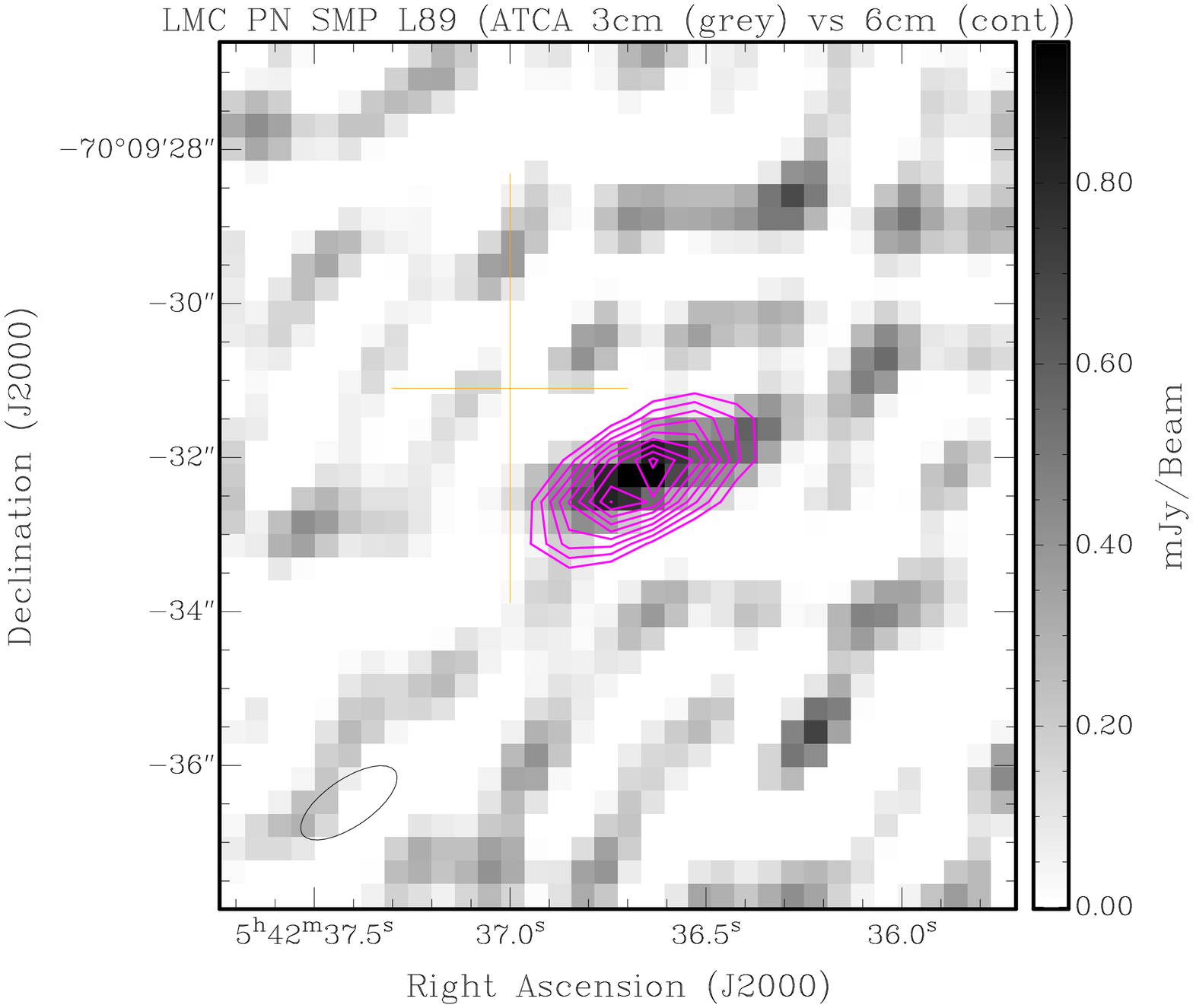}
\includegraphics[angle=-90, trim=0 0 0 0, width=.475\textwidth]{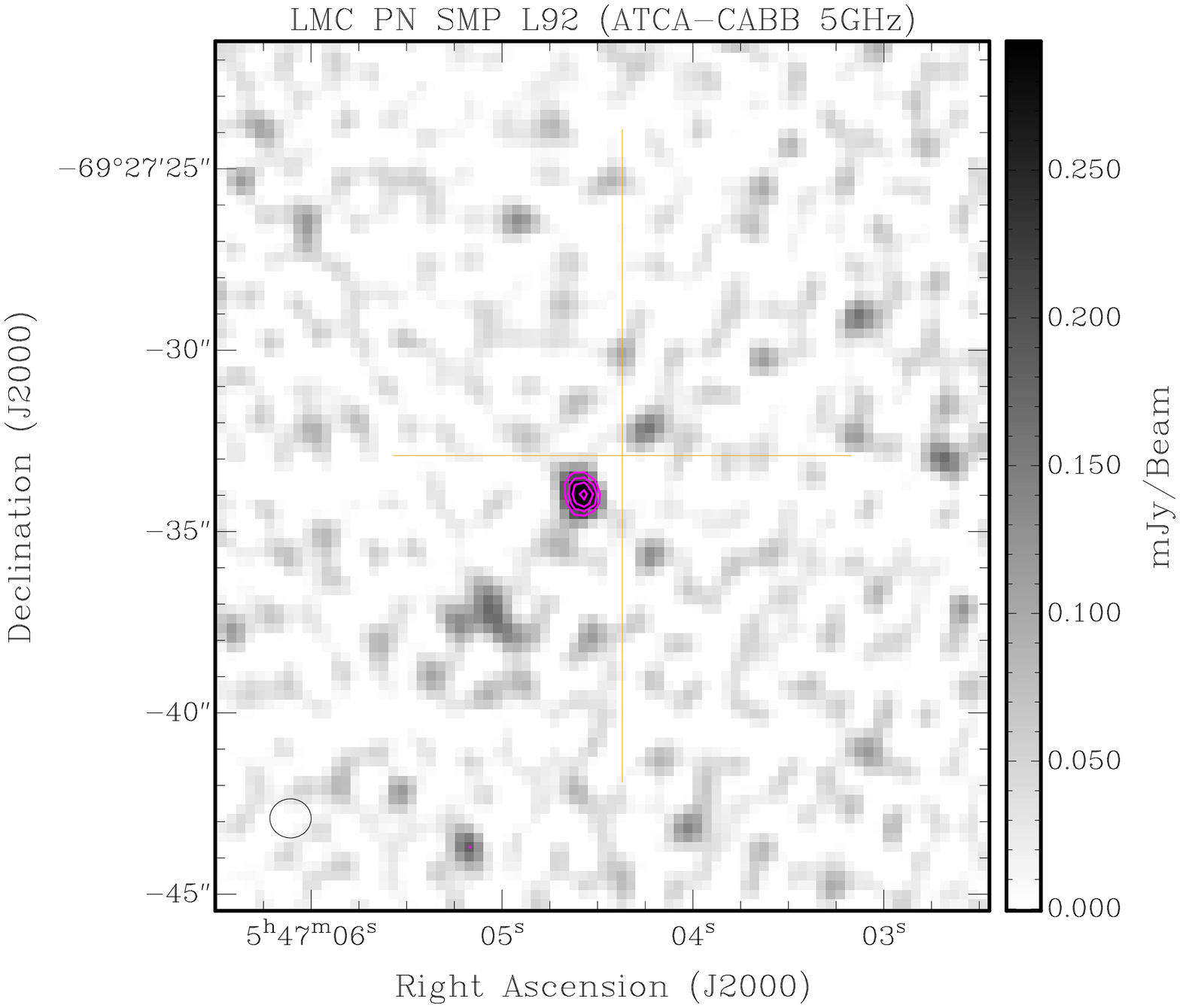}
\caption{New and previous radio detections of LMC PNe. Images are constructed as total intensity radio maps. Contours are integral multiples of the measured RMS noise starting at 3$\sigma$ with spacings of 1$\sigma$. Top row: SMP\,L84 at 6~cm and SMP\,L85 at 5~GHz; Bottom row: SMP\,L89 at 3~cm (grey scale) with 6~cm contours and SMP\,L92 at 5~GHz. The beam size of each image is shown in the bottom left corner. The orange crosses represent the PN positions from the RP catalogue.}
\label{fig:radio_gr5}
\end{figure*}

PNe that are resolved at any observational wavelength are frequently found to have an irregular shape \citep{Kwok2010}. This is easily seen in high resolution images of PNe at optical wavelengths and in some of our resolved RC images: SMP~L13, SMP~L21, SMP~L37, SMP~L38, SMP~L58, for example (see Fig.~\ref{fig:radio_gr1}--\ref{fig:radio_gr5}). As the dimensions of the synthesised beam approaches the overall size of the PN, the detected shape becomes more indistinct approaching a Gaussian intensity distribution defined only by the source intensity and the synthesised beam parameters. For unresolved PNe, the coordinates can be approximated as the position of the peak of a Gaussian-fitting to the intensity. For resolved images, however, there is no consistent portion of the typically asymmetric image to unambiguously define as its coordinates since only a portion of the PN may be detected. The RP catalogue positions are in the visible portion of the spectrum which may image different parts of the PN from the RC data. The position of the central star (CS) is likely the best definition if it can be detected but RC imaging does not detect the CS.

The synthesised telescope beam for each radio detection is shown in the images of the PNe \mbox{(Fig.~\ref{fig:radio_gr1}--\ref{fig:radio_gr5}).} The synthesised beam is a two-dimensional Gaussian distribution dependent on the wavelength, the ATCA telescope array configuration, and the coordinates of the image. The optical diameters of the PNe we studied here range over an order of magnitude -- from 0.24~arcsec to 2.53~arcsec. The synthesized beam sizes are typically not symmetric and vary widely as seen in Table~\ref{tbl:data} (Col.~4) from 0.7~arcsec to 10.0~arcsec. In addition to meeting the requirement that the intensity measured is $\geqslant$3$\sigma$ above the RMS noise, we require that a PN RC detection meets one of two additional conditions: 1) for unresolved sources the peak of the RC intensity must be $<$5~arcsec from the RP PN coordinates; or 2) the apparent centre of a resolved source must be $<$5~arcsec from the RP PN coordinates.  We have chosen  5~arcsec since the maximum absolute positional uncertainty of the ATCA is 5~arcsec \citep{Stevens2014}.

%

%

%

%

%

 \subsection{Detection Results} \label{sec:detresults}

After careful examination of all of our available data, 21 new RC detections of PNe in our base catalogue were identified: SMP\,L13, SMP\,L15, SMP\,L21, SMP\,L23, SMP\,L29, SMP\,L37, SMP\,L38, SMP\,L45, SMP\,L50, SMP\,L52, SMP\,L53, SMP\,L58, SMP\,L63, SMP\,L66, SMP\,L73, SMP\,L75, SMP\,L76, SMP\,L78, RP\,659, SMP\,L85, and SMP\,L92. 

For seven previously detected PNe \citep{Filipovic2009}: SMP\,L47, SMP\,L48, SMP\,L62, SMP\,L74, SMP\,L83, SMP\,L84, and SMP\,L89, a flux density was measured on at least one frequency, confirming previous detections. The data for these PN detections come from the various ATCA projects listed in Table~\ref{tbl:data}\footnote{Also refer to \cite{Shaw2006} for a detailed description of optical characteristics of many of these PNe.}. All of the RC LMC PNe detected here are from the set of PN coordinates that RP has published with a classification of `Known' or `True'.

The four PNe listed in \cite{Filipovic2009} for which we did not find any new reliable detections were SMP\,L8, SMP\,L25, SMP\,L33, and SMP\,L39. SMP\,L8 has been reclassified as a compact \HII\ region and is not listed in the RP catalogue. Therefore, they are not included in our study presented here. We detected RC emission from the positions of previously known RC PNe SMP\,L25 and SMP\,L33 in 20~cm data from ATCA project C354. Unfortunately, both PNe are far from the centre of the primary beam ($\sim$25~arcmin at 20~cm) and therefore their flux density measurements are very uncertain even after the image has been corrected for the primary beam ({\sc miriad} {\sc linmos} task). SMP\,L39 was not contained in any of the projects which met our requirements for inclusion. 

The RP coordinates and the measured integrated radio flux densities for the detected PNe are shown in Table~\ref{tbl:MeasureRC}. The flux densities of the three previously detected PNe  which we could not measure, SMP\,L25, SMP\,L33, SMP\,L39, are included in Table~\ref{tbl:MeasureRC} with intensities from \cite{Filipovic2009} for completeness of the LMC PN sample. The finding charts for all LMC PN RC detections are shown in Figs.~\ref{fig:radio_gr1} through \ref{fig:radio_gr5}. These images are constructed as total intensity RC maps at the detection frequencies overlaid with contours at integral multiples of the estimated RMS noise. The orange crosses represent the PN coordinates from the RP catalogue. 

As can be seen from the finding charts presented \mbox{(Fig.~\ref{fig:radio_gr1}--\ref{fig:radio_gr5})}, in most cases the radio detections are of isolated sources well correlated with their RP catalogue optical counterparts.  Most detections of a PN were at one or two frequencies with only two detections at three frequencies and none at four frequencies. The absence of detection at more frequencies was due either to a lack of data for a PN at that frequency or that the emissions were determined to be below the 3$\sigma$ noise level.

\begin{landscape}
\begin{table} 
\centering
\scriptsize
\caption{Coordinates and flux density measurements of detected LMC PNe. The new flux density measurements are shown in the columns titled `new' and previous detections \citep{Filipovic2009} are listed in the columns titled `F09'. The symbol `$<$' indicates that the actual value is less than the indicated value. Col.~12 is the MIR/S$_{20\,cm}$ ratio using $8~\umu$m data from the VizieR Online Data Catalog: SAGE LMC and SMC IRAC Source Catalog (IPAC 2009). Col.~13 is the calculated \Halpha\ flux using tabulated \HB\ and extinction values c\HB\ from \citep{Reid2010}. Col.~14 is from Table~\ref{tbl:physprop}. The `Source Ref.', Col~15, refers to the sources of the RC data used for our measurements from the ATCA project data base references found at the bottom of this table.}  
 \begin{tabular}{rlcccccccccccccl}
 \hline
No.	&PN			&RA 			&DEC 		&\multicolumn{2}{c}{S$_{3\,cm}$~(mJy)}			&\multicolumn{2}{c}{S$_{6\,cm}$~(mJy)}&\multicolumn{2}{c}{S$_{13\,cm}$~(mJy)}&	\multicolumn{2}{c}{S$_{20\,cm}$~(mJy)} &MIR/ & \Halpha~flux $(10^{-13})$ & $\theta$  & Source\\
	& 			&(J2000)		&	(J2000)			&new				&F09		&new				&F09			&new				&F09		&new				&F09 &S$_{20\,cm}$&${\text{erg c}}{{\text{m}}^{ - 2}}{\text{ }}{{\text{s}}^{ - 1}}$ & (\arcsec)&   Ref.\\
    &(1)  		&(2)				&(3)					&(4)				& (5)	&(6)				&(7)			&(8)				&(9)		&(10)			&(11) &(12) &(13) &(14)  &(15)   \\ 
  \hline
1	&SMP\,L13	&04:59:59.99	&--70:27:40.8		&...				&...				&0.40$\pm$0.04	&...				&...				&...		&0.45$\pm$0.04	&...				&18.0 &5.8 &0.94 & 8 \\
2	&SMP\,L15	&05:00:52.71	&--70:13:41.1		&...				&...				&0.64$\pm$0.05	&...				&0.98$\pm$0.05	&...		&0.90$\pm$0.06	&...				&24.4 &6.5 &0.82 & 4,8 \\
3	&SMP\,L21	&05:04:51.99	&--68:39:09.7		&...				&...				&0.62$\pm$0.05	&...				&...				&...		&...				&...				&... &6.9 &0.24 & 8 \\
4	&SMP\,L23	&05:06:09.45	&--67:45:27.8		&...				&...				&...				&...				&0.67$\pm$0.04	&...		&0.66$\pm$0.04	&...				&4.4 &6.7 &0.39 & 5 \\
\smallskip
5	&SMP\,L25	&05:06:23.91	&--69:03:19.0		&...				&2.4$\pm$0.24	&...				&2.1	$\pm$0.21	&...				&...		&...				&1.8$\pm$0.2		&10.5 &14.7 &0.46 &... \\
6	&SMP\,L29	&05:08:03.32	&--68:40:16.6		&...				&...				&0.65$\pm$0.06 	&...				&...				&...		&0.70$\pm$0.07	&...				&17.9 &5.9 &0.54 & 3,8 \\
7	&SMP\,L33	&05:10:09.41	&--68:29:54.5		&...				&1.8$\pm$0.2		&...				&$<$1.5			&...				&...		&...				&$<$1			&... &5.1 &0.62 & ...  \\
8	&SMP\,L37	&05:11:02.92	&--67:47:59.4		&...				&...				&0.52$\pm$0.04	&...				&...				&...		&...				&...				&... &3.3 &0.50 & 8  \\
9 	&SMP\,L38	&05:11:23.69	&--70:01:57.4		&...				&...				&0.58$\pm$0.05 	&...				&...     		&...		&0.58$\pm$0.05	&...				&87.9 &7.0 &0.50 & 6,8  \\
\smallskip
10	&SMP\,L39	&05:11:42.11 &--68:34:59.1		&...				&$<$1.5			&...				&$<$1.5			&...				&...		&...				&2.3	$\pm$0.23	&3.0 &2.2 &0.46 &...  \\		
11	&SMP\,L45	&05:19:20.71	&--66:58:06.9		&...				&...				&...				&...				&1.00$\pm$0.10	&...		&...				&...				&... &3.6 &1.44 & 9  \\
12	&SMP\,L47	&05:19:54.65	&--69:31:05.1		&...				&2.3$\pm$0.23	&...				&2.1$\pm$0.21	&1.50$\pm$0.05	&...		&1.30$\pm$0.05	&2.2$\pm$0.22		&17.1 &11.3 &0.50 & 1 \\
13	&SMP\,L48	&05:20:09.48	&--69:53:39.1		&...				&$<$1			&1.4$\pm$0.2		&1.5$\pm$0.2		&1.50$\pm$0.10	&...		&1.00$\pm$0.05	&1.6$\pm$0.2		&18.0 &10.2 & 0.40& 1  \\
14	&SMP\,L50	&05:20:51.73	&--67:05:43.4		&...				&...				&...				&...				&1.00$\pm$0.15	&...		&...				&...				&... &6.5 &0.70 & 9  \\
\smallskip
15  &SMP\,L52	&05:21:23.96	&--68:35:33.8		&...				&...				&1.00$\pm$0.05	&...				&...				&...		&...				&...				&... &11.1 &0.40 & 8 \\
16	&SMP\,L53	&05:21:32.83	&--67:00:04.6		&...				&...				&0.61$\pm$0.06	&...				&0.92$\pm$0.15	&...		&...				&...				&... &20.7 &0.80 & 8,9 \\
17  &SMP\,L58	&05:24:20.77	&--70:05:01.5		&...				&...				&0.31$\pm$0.05  	&..				&...				&...		&...				&...				&... &9.0 &0.26 & 8 \\
18	&SMP\,L62	&05:24:55.08	&--71:32:56.3		&2.1$\pm$0.2	    &1.8$\pm$0.2		&2.6$\pm$0.1 	&2.1$\pm$0.21	&...				&...		&...				&2.5	$\pm$0.3		&3.0 &18.8 &0.54 & 2  \\
19  &SMP\,L63	&05:25:26.02	&--68:55:54.0		&0.5$\pm$0.1 	&...				&1.2$\pm$0.1		&...				&...				&...		&...				&...				&... &12.7 &0.52 & 8 \\
\smallskip
20  &SMP\,L66	&05:28:41.00	&--67:33:39.3		&...				&...				&0.25$\pm$0.05	&...				&...				&...		&...				&...				&... &8.3 &1.08 & 8  \\
21  &SMP\,L73	&05:31:21.89	&--70:40:45.5		&...				&...				&1.01$\pm$0.06	&...				&...				&...		&...				&...				&... &13.7 &0.50 & 8 \\
22  &SMP\,L74	&05:33:29.76	&--71:52:28.5		&0.97$\pm$0.15	&$<$1.5			&1.06$\pm$0.15	&$<$1.5			&...				&...		&...				&3.0$\pm$0.3		&5.2 &... &0.70 & 8 \\
23  &SMP\,L75	&05:33:47.00	&--68:36:44.2		&1.14$\pm$0.15	&...				&1.45$\pm$0.15	&...				&...				&...		&...				&...				&... &10.2 &0.40 & 8 \\
24  &SMP\,L76	&05:33:56.10	&--67:53:09.2		&...         	&...				&0.61$\pm$0.07	&...				&...				&...		&...				&...				&... &13.7 &... & 8 \\
\smallskip
25  &SMP\,L78	&05:34:21.19	&--68:58:25.3		&...				&...				&1.28$\pm$0.06	&...				&...				&...		&...				&...				&... &12.5 &0.56 & 8 \\
26	&RP\,659		&05:35:10.19	&--69 39 39.3		&...				&...				&0.12$\pm$0.02	&...				&...				&...		&...				&...				&... &1.2 &... & 7 \\
27	&SMP\,L83	&05:36:20.81	&--67:18:07.5		&...				&0.7$\pm$0.07	&0.74$\pm$0.10	&0.9$\pm$0.1		&...     		&...		&0.9$\pm$0.1		&1.6	$\pm$0.2		&1.4 &4.9 &2.52 & 3  \\
28  &SMP\,L84	&05:36:53.46	&--71:53:39.3		&...				&$<$1.5			&0.80$\pm$0.11 	&$<$1.5			&...				&...		&...				&1.6$\pm$0.2		&1.7 &... &0.58 & 8 \\
29  &SMP\,L85	&05:40:30.80	&--66:17:37.4		&...				&...				&1.15$\pm$0.06 	&...				&...				&...		&...				&...				&... &6.5 &... & 8 \\
\smallskip
30  &SMP\,L89	&05:42:37.17	&--70:09:30.4		&1.60$\pm$0.15	&1.8$\pm$0.2		&1.49$\pm$0.10	&1.6$\pm$0.2		&...				&...		&...				&2.1$\pm$0.21	&13.3 &7.9 &0.46 & 8 \\
31  &SMP\,L92	&05:47:04.80	&--69:27:32.1		&...				&...				&0.41$\pm$0.06	&...				&...				&...		&...				&...				&... &6.6 &0.52 & 8  \\
 \hline
 \end{tabular}
 \smallskip
 \flushleft
ATCA project numbers as in Table~\ref{tbl:data} - 1:~C256, 2:~C308, 3:~C354, 4:~C395, 5:~C479, 6:~C520, 7:~C1973, 8:~C2367, 9:~C2908.
\label{tbl:MeasureRC}
\end{table}\end{landscape}

\begin{table*}
\centering
\normalsize
\caption{Measured and calculated LMC PN RC parameters. Col.~2 contains the photometric angular diameters. The references used for the diameters are listed in Col.~3 and they refer to the list at the bottom of the table. The spectral indexes ($\alpha$) are listed in Col.~4a and 4b. The adopted flux densities at 6~cm are in Col.~5. Those which are in brackets were estimated from other band measurements. The angular diameters in Col.~2 in arcseconds are used to calculate the radii in pc shown in Col.~7. Values for brightness temperatures (T$_b$), the electron densities (n$_e$) and ionised masses (M$_i$) are calculated from the adopted 6~cm flux densities and the radii. These are presented in Cols.~6, 8, and 9, respectively.	}  
 \begin{tabular}{lcccccccccccc}
 \hline
 \hline							
PN			&$\theta$	&Ref    & \multicolumn{2}{c} {$\alpha$} &S$_{6\,cm}$		&T$_b$	&r		&log($n_e$)			&$M_i$\\
			&(\arcsec)	&		&$>$5~GHz		&$<$~5~GHz		&(mJy)			&(K)		&(pc)	&	(cm$^{-3}$)			&($M_{\odot}$)\\
(1)			&(2)		&(3)		&  (4a)		& (4b)				&(5)		&(6)		&(7)		&(8)		&(9)\\
 \hline
SMP\,L13		&0.94	&1		&...				&--0.09$\pm$0.10  	&0.40$\pm$0.04			&25$\pm$3		&0.11	&3.19	&0.225 \\
SMP\,L15		&0.82	&1		&...				&--0.27$\pm$0.10  	&0.64$\pm$0.05			&68$\pm$5		&0.10	&3.38	&0.236\\
SMP\,L21		&0.24	&4		&...				&...					&0.62$\pm$0.05			&594$\pm$48		&0.03	&4.17	&0.036\\
SMP\,L23		&0.40	&3		&...				&\p00.28$\pm$0.20 	&$[$0.68$\pm$0.04$]$		&302$\pm$18		&0.05	&3.87	&0.075\\
\smallskip
SMP\,L25		&0.46	&1		&\p00.27$\pm$0.40	&\p00.11$\pm$0.20		&2.1$\pm$0.21	&559$\pm$56		&0.06	&4.02	&0.167\\
SMP\,L29		&0.54	&2		&...				&--0.05$\pm$0.13		&0.65$\pm$0.06			&124$\pm$11		&0.07	&3.65	&0.128\\
SMP\,L33		&0.62	&1		&...				&...					&$[$1.9$\pm$0.2$]$		&339$\pm$36		&0.07	&3.80	&0.252\\
SMP\,L37		&0.50	&2		&...				&...					&0.52$\pm$0.04			&116$\pm$9		&0.06	&3.66	&0.096\\
SMP\,L38		&0.50	&1		&...				&\p00.00$\pm$0.13		&0.58$\pm$0.05			&130$\pm$11	&0.06	&3.68	&0.103\\
\smallskip
SMP\,L39		&0.46	&2		&...				&...					&$[$2.0$\pm$0.2$]$		&645$\pm$65		&0.06	&4.00	&0.168\\
SMP\,L45		&1.44	&2		&...				&...					&$[$0.88$\pm$0.09$]$		&29$\pm$3		&0.17	&3.09	&0.612 \\
SMP\,L47		&0.50	&2		&...				&\p00.26$\pm$0.13		&2.1$\pm$0.21		&493$\pm$49		&0.06	&3.96	&0.203\\
SMP\,L48		&0.40	&2		&...				&\p00.22$\pm$0.08		&1.4$\pm$0.2			&596$\pm$85		&0.05	&4.02	&0.116\\
SMP\,L50		&0.70	&1		&...				&...					&$[$0.93$\pm$0.14$]$		&128$\pm$19		&0.08	&3.56	&0.216 \\
\smallskip
SMP\,L52		& 0.40	&2		&...				&...					&1.0$\pm$0.05			&349$\pm$17		&0.05	&3.94	&0.096 \\
SMP\,L53		& 0.80		&1		&...				&--0.44$\pm$0.30  	&0.61$\pm$0.06		&53$\pm$5		&0.10	&3.39	&0.211 \\
SMP\,L58		&0.26	&1		&...				&...					&0.31$\pm$0.05			&260$\pm$42		&0.03	&3.98	&0.028\\
SMP\,L62		&0.54	&2		&--0.43$\pm$0.30	&...					&2.6$\pm$0.1				&496$\pm$19		&0.07	&3.95	&0.251\\
SMP\,L63		&0.52	&2		&--1.80$\pm$0.60		&...				&1.2$\pm$0.1				&251$\pm$21		&0.06	&3.84	&0.160\\
\smallskip
SMP\,L66		& 1.10	&4		&...				&...					&0.25$\pm$0.05			&12$\pm$2		&0.13	&2.99	&0.214 \\
SMP\,L73		& 0.50	&2		&...				&...					&1.01$\pm$0.06			&226$\pm$13		&0.06	&3.80	&0.135 \\
SMP\,L74		& 0.70	&2		&--0.18$\pm$0.60	&...					&1.06$\pm$0.15			&121$\pm$17		&0.08	&3.59	&0.236\\
SMP\,L75		& 0.40	&2		&--0.50$\pm$0.50		&...				&1.45$\pm$0.15			&507$\pm$52		&0.05	&4.03	&0.115\\
SMP\,L76		&...		&...		&...				&...					&0.61$\pm$0.07			&...		&...		&...		&...	\\
\smallskip
SMP\,L78		& 0.56	&1		&...				&...					&1.28$\pm$0.06			&230$\pm$11		&0.07	&3.78	&0.179\\
RP\,659		&...		&...		&...				&...					&0.12$\pm$0.02			&...		&....	&....	&...  \\
SMP\,L83		& 2.53  &2		&...				&--0.14$\pm$0.20  	&0.74$\pm$0.10			&7$\pm$1			&0.30	&2.68	&1.337\\
SMP\,L84		& 0.50		&2		&...				&...					&0.80$\pm$0.11		&132$\pm$18		&0.07	&3.65	&0.150\\
SMP\,L85		&...		&...		&...				&...					&1.15$\pm$0.06			&...		&...		&...		&...	 \\
\smallskip
SMP\,L89		&0.46	&2		&\p00.14$\pm$0.30	&...					&1.49$\pm$0.10		&397$\pm$27		&0.06	&3.94	&0.143 \\
SMP\,L92		&0.52	&2		&...				&...					&0.41$\pm$0.06			&86$\pm$13		&0.06	&3.58	&0.090 \\
  \hline
\end{tabular}
 \smallskip
 \flushleft
1:~\cite{Shaw2001}, 2:~\cite{Shaw2006}, 3:~\cite{1998ApJ...503..253V}, 4:~\cite{Stanghellini1999}.
 \label{tbl:physprop}
\end{table*}

\subsection{Radio-continuum properties of the detected PNe} \label{sec:radprop}

From the measurements made here and data from the literature, additional physical properties of the detected PNe were determined. In Table~\ref{tbl:MeasureRC} (Column~12) we calculated the ratio of Mid-Infrared (MIR) and 20~cm flux densities using $8~\umu$m MIR data from the Sage LMC and SMC IRAC Source Catalog (IPAC 2009)\footnote{http://adsabs.harvard.edu/abs/2012yCat.2305....0G } containing LMC data from \cite{Meixner2006}. This data was downloaded from the VizieR Online Data Catalog collection. \citet{Cohen2007a} using Galactic data from \citet{Cohen2001} and other sources referenced therein, examined the MIR-RC ratio, ${{{S_{8.3\mu m}}} \mathord{\left/ {\vphantom {{{S_{8.3\mu m}}} {{S_{843\,MHz}}}}} \right. \kern-\nulldelimiterspace} {{S_{\rm 843\,MHz}}}}$ for PNe in the Galactic plane and calculated a median ratio of 12. The median ratio for our population of the LMC PNe is 11.9 using ${{{S_{8\mu m}}} \mathord{\left/ {\vphantom {{{S_{8\,\mu m}}} {{S_{1.4\,GHz}}}}} \right. \kern-\nulldelimiterspace} {{S_{\rm 1.4\,GHz}}}}$. The largest deviation from this median in our sample was calculated to be 87.9 for SMP\,L38. This, however, is likely to be due to unusual dust emissions from polycyclic aromatic hydrocarbons emitted by this PN as discussed in \cite{Bernard-Salas2009} who included this PN in their study of the MC PNe. 

In Table~\ref{tbl:physprop}, the RC spectral energy distribution (SED) was calculated for each PN for which measurements of flux density had been made on at least two frequencies. The fitting equation is:

\begin{equation}
 S_{\nu}= \beta\nu^{\alpha} 
\label{SED_equation} 
\end{equation}

\noindent where $\alpha$ is the spectral index and $S_\nu$ is the flux density at frequency $\nu$. The spectral indices are shown in Cols.~4a and 4b of Table~\ref{tbl:physprop}. 

PN RC emission is primarily thermal radiation from free-free interactions between electrons and ions within the nebular shell \citep{Kwok2000}. The SED is dependent on the metallicity, electron density, and temperature. Typically, at higher frequencies ($>$5~GHz), the nebula is optically thin and the spectral index ($\alpha$) is expected to be $\sim -0.1$. At lower frequencies ($<$5~GHz), young PNe are most likely optically thick with an expected spectral index of $\sim$2 \citep{Pottasch1984}. However, different density profile models can produce a very different SED.  For example, if the density profile is from a constant mass loss, the spectral index for the PN would be optically thin and observed to be inverted with $\alpha\sim$0.6 \citep{Panagia1975}. \cite{Gruenwald2007} cautions that radio SED observations from PNe cannot uniquely determine the density profiles of the PNe. 

The critical frequency of a PN is defined as the transition frequency from optically thick with a positive spectral index to optically thin with a negative spectral index. Most Galactic PNe exhibit a critical frequency of less than 5~GHz \citep{Gruenwald2007}. However, the critical frequencies for the PNe depend on gas density and can range from $\sim$0.4~GHz to $>$15~GHz \citep{Gruenwald2007}. Young and compact PNe have been observed with critical frequencies of $\sim$5~GHz to 28~GHz \citep{Aaquist1991}. These young PNe most likely have very high electron densities which makes the critical frequency much higher than 5~GHz \citep{Kwok1982,Pazderska2009}. We calculated the spectral indices $\alpha$ above and below 5~GHz from Eq.~\ref{SED_equation} as a proxy of the evolutionary state of the PNe.

Non-thermal radio emission has been proposed for some PN emissions \citep{Dgani1998}. The mechanism offered requires fast winds from the CS of the PN to create an inner region where strong magnetic fields are interacting with the CS wind. A non-thermal SED would be expected to have a spectral index $\alpha\ll-0.1$ \citep{GurzadyanGrigorA.1997}. \cite{Cohen2006} reported non-thermal emissions from a very unusual Asymptotic Giant Branch (AGB) star. This star was measured to have $\alpha\sim-0.9$ and contained shocked interactions between a cold PN--like nebular shell and the hot, fast stellar wind from the AGB star. The source, however, of most apparent non-thermal radio detections is more likely to be a Supernova Remnant or an Active Galactic Nucleus coincident with the position of the PN.

In Col.~5 (Table~\ref{tbl:physprop}) the adopted flux densities at 6~cm for all of the detected RC PNe are presented. If direct measurement at 6~cm was not available, we estimated the flux density from neighbouring bands. This was necessary for SMP~L23, SMP~L33, SMP~L39, and SMP~L45. The entries that are estimated are indicated in Table~\ref{tbl:physprop} with brackets around the 6~cm values.

\begin{figure*}
\includegraphics[width=0.75\textwidth]{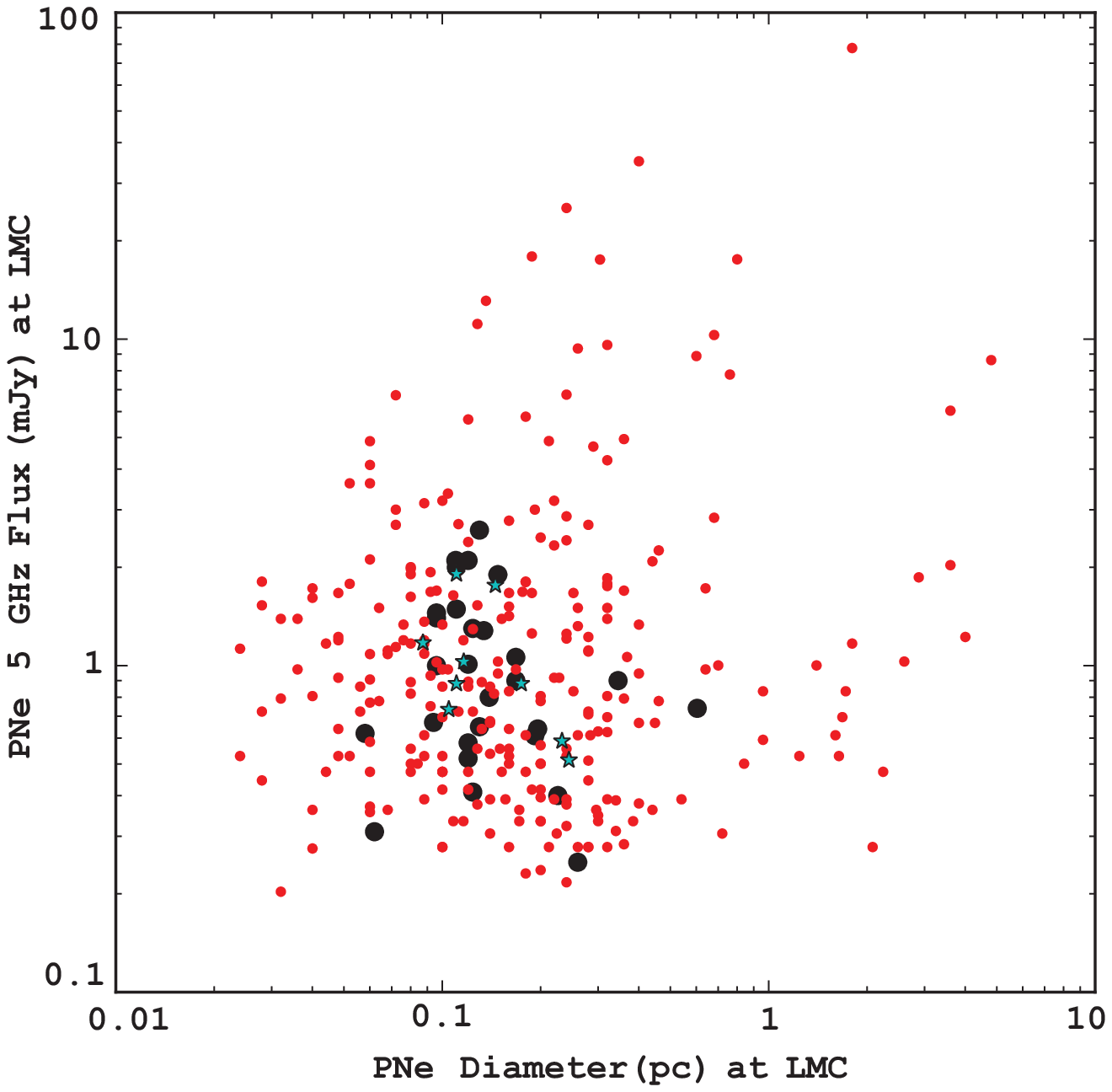}
\caption{Galactic PNe from \citep{Siodmiak2001} (small red dots), SMC PNe data from \citep{leverenz2016} (green stars), (both scaled to the distance of the LMC), and our new LMC measurements (large black dots). }
\label{fig:compars}
\end{figure*}

We also compared our adopted 6~cm flux densities with a sample of well-measured Galactic PNe (GPNe). This sample is a data base of optical diameters and 6~cm integrated flux density measurements from \cite{Siodmiak2001} consisting primarily of PNe from the Galactic bulge. Their diameter measurements were calculated using the 10 per~cent~peak contour approach. They used only optical data from photographic plates since they judged that newer CCD images were too sensitive and could not be directly compared to extant diameter measurements derived from photographic plates. We re-scaled their data to the distance of the LMC (50~kpc; \citet{Pietrzynski2013}) and plotted both data-sets on the same figure (Fig.~\ref{fig:compars}). Additionally we included nine SMC PNe from \cite{leverenz2016}: SMP~S6, SMP~S13, SMP~S14, SMP~S16, SMP~S17, SMP~S18, SMP~S19, SMP~S22, and SMP~S24. These SMC PNe were unambiguously determined not to be PN mimics and were also re-scaled to the distance of the LMC. The green star markers on Fig.~\ref{fig:compars} represent the SMC data.

Our RC LMC PN measurements are consistent with the scaled flux densities and diameters from both the Galactic and SMC samples. The angular diameters in Table~\ref{tbl:physprop} (Col.~2) were adopted from \citet{Shaw2001,Shaw2006}, \citet{1998ApJ...503..253V}, and \citet{Stanghellini1999}. For consistency, we chose only photometric radii defined as the radius containing 85 per cent of the total emitted flux \citep{Shaw2001} since measurements based on this  method were available from all of the sources we consulted for the radius data which we adopted. We estimated that the differences in the techniques for the diameter measurements for this comparison was not significant (the 10 per cent peak flux density used by \cite{Siodmiak2001} for the GPNe compared to the 85 per cent flux density limit we employed for the MC PNe).

We used our adopted 6~cm flux densities and the angular diameters from the literature to estimate the electron densities ($n_e$) and ionized masses ($M_i$) for all detected RC PNe (Table~\ref{tbl:physprop}, Cols.~8 and 9). The $n_e$ and $M_i$ are calculated from equations (1) and (2) in \cite{gathier1987}, respectively. The electron temperature ($T_{e}$) for all PNe was chosen to be the canonical temperature of 10\,000~K. The solar metallicity parameter and the consensus filling factors were both assigned a numerical value of $0.3$ \citep{Boffi1994,Frew2016a}. The helium abundance for SMP\,L13 was taken from \cite{DeFreitasPacheco1993}. SMP\,L39, SMP\,L52, and RP\,659 were assigned the default value of 0.1 per \cite{gathier1987} since measured values for those PNe were not available. Calculations for the remaining PNe used helium abundances from \cite{Chiappini2009}. No values for the fraction of doubly ionised helium were available so we adopted the default value of $ 1/3 $ as used by \cite{gathier1987}. Uncertainties in $n_e$ and $ M_i $ are estimated to be on the order of 40~per cent. 

Thermal brightness (T$_b$) at 6~cm is shown in Table~\ref{tbl:physprop} (Col.~6). It is a good proxy for PN evolutionary status \citep{Kwok1985,zijlstra1990,Gruenwald2007} and also an indication of self absorption. We calculated the conventional radio thermal brightness assuming an isothermal nebula from \cite{Wilson2009} as follows: 
\begin{equation}
{T_b} = \frac{{{c^2}}}{{2k{\nu ^2}}} \cdot \frac{{{S_\nu }}}{{\Delta \Omega }}
\label{T_sub_b_equation} 
\end{equation}
\noindent To determine the self absorption correction for our results, we applied the two component model of self absorption from \cite{Siodmiak2001}. We adopted the values of \mbox{$\xi=0.27$} and $\varepsilon=0.19$ for this comparison. This model provides a correction factor based on the ratio of the measured $ T_{b} $ with the electron temperature, $ T_{e} $. We adopted the canonical value of 10000~K for $T_{e}$. The highest thermal brightness we calculated from our data is from SMP\,L39 at 645~K. For this value of $ T_{b} $, the model gives a correction factor of 1.8 at 20~cm and 1.01 at 6~cm. The value of 339~K for $ T_{b}$ from SMP\,L33 is determined to have a correction factor of 1.4 at 20~cm. At 6~cm, there is essentially no self absorption. Our conclusion was that there was significant self absorption in some of our results at 20~cm but at shorter wavelengths the self absorption was negligible.

\section{$\Sigma- D$ relation for PNe}

\begin{table*}
\caption{Results of the $\Sigma$--$D$ analysis at 6~cm. Col.~1 lists the sample designation with the number of data points in Col.~2 and the calculated correlation coefficient in Col.~3. Cols.~4-7 list the values for the parameters of the fitting from the $\Sigma$ offsets and their estimated uncertainties. The average fractional error for the calculated distances are in Col.~8. Cols.~9-13 present the values for the orthogonal-offset-fitting and the corresponding average distance fractional errors. In Cols.~14-16 we present the average fractional errors for the PDF distance estimators: mean, mode, and median.}
\scriptsize
\begin{tabular}{cccccccccccccccccccccc}\hline\hline
Sample&&&\multicolumn{5}{c}{$\Sigma$ offsets}&&\multicolumn{5}{c}{Orthogonal offsets}&&\multicolumn{3}{c}{$f(\%)$}\\\cline{4-8}\cline{10-14}\cline{16-18}
&$N$&$r$&$A$ & $\Delta A$ & $\beta$ & $\Delta \beta$&$f(\%)$ && $A$ & $\Delta A$&$\beta$ & $\Delta \beta$&$f(\%)$ && mean & mode & median\\
(1)&(2)&(3)&(4)&(5)&(6)&(7)&(8)&&(9)&(10)&(11)&(12)&(13)&&(14)&(15)&(16)\\\hline
LMC&28&--0.87&--19.7 &0.3&2.2&0.4&22.97&&--20.3& 0.4&2.9&0.5&19.58&&18.29&17.28&18.37\\
SMC$^\mathrm{\dagger}$&9&--0.94&--20.1&...&2.7&...&9.99&&--20.4&...&3.0&...&9.52&&...&...&...\\
LMC+SMC&37&--0.88&--19.9&0.3&2.3&0.3&20.17&&--20.3&0.3&2.9&0.4&17.00&&16.05&14.64&16.02\\\hline
\end{tabular}
\\$^\dagger$ Presented results are not reliable due to the small number of data points and are given only for the sake of completeness. Unlike the other two samples, these fitting parameters are not determined with the bootstrap procedure described in the paper but from a simple linear regression fitting to the data sample. The 2D density smoothing did not yield convergent results and no fractional errors were calculated. 
\label{SigmaD_results_table}
\end{table*}

The radio flux measurements at 6~cm and radius measurements in parsecs from Table~\ref{tbl:physprop} were used to calculate the surface brightness and diameter values for the $\Sigma$--$D$ relation study of the MC PNe.

Recently, \citet{leverenz2016} described the $\Sigma$--$D$ relation for a sample of SMC PNe. Here, we extended that work by analyzing our radio detected PNe in terms of the $\Sigma$--$D$ relation for both the LMC and the SMC. The model for the radio surface brightness we adopted from \cite{Vukotic2009} is shown in the following equation: 

\begin{equation}
\Sigma \left[ {\text{W}}{{\text{m}}^{ - 2}}{\text{H}}{{\text{z}}^{ - 1}}{\text{s}}{{\text{r}}^{ - 1}} \right] = 1.505 \cdot {10^{ - 19}}{{S\left[ \text{Jy} \right]} \mathord{\left/
 {\vphantom {{S\left[ {Jy} \right]} {{\theta ^2}\left[ ' \right]}}} \right.
 \kern-\nulldelimiterspace} {{\theta ^2}\left[ \arcmin \right]}}
\end{equation}

\noindent The form of the $\Sigma$--$D$ model equation adopted for these surface brightness calculations is

\begin{equation}
\Sigma = AD^{-\beta}
\label{SigD_equation}
\end{equation}

\noindent see \citet{Vukotic2014}.

There are 715 observed PNe in the LMC \citep{Reid2013,Reid2014} but only 31 of them have been detected at radio frequencies. This is indicative of a strong observational bias suggesting that only the `tip of the iceberg' is observed at radio frequencies, see \cite{Vukotic2009}. Indeed, while the measured surface brightness of GPNe extends from $\sim10^{-24}$ to $\sim10^{-16}$ ${\text{W}}{{\text{m}}^{ - 2}}{\text{H}}{{\text{z}}^{ - 1}}{\text{s}}{{\text{r}}^{ - 1}}$ , the PNe from the MCs are detected only in the high brightness half of this interval from $\sim 10^{-20}$ to $\sim 10^{-16}$ ${\text{W}}{{\text{m}}^{ - 2}}{\text{H}}{{\text{z}}^{ - 1}}{\text{s}}{{\text{r}}^{ - 1}}$. Similar selection effects can be seen in $D$ as well. The largest PN in our radio LMC sample is smaller than 1 pc, unlike the largest PN used for {\Halpha} statistical distance calibration \citep{Frew2016a} being $>2$ pc in size. 

With such a strong selection effect in play we constructed a Monte Carlo simulation (described in Section \ref{sensitivity}) to quantify the influence of sensitivity related selection effects on our $\Sigma$--$D$ samples. We utilised bootstrap statistics \citep{EfronTibshirani93} to reconstruct the prior distribution of the $\Sigma$--$D$ relation parameters. For each $\Sigma$--$D$ data sample we applied both $\Sigma$ offset and orthogonal offset analysis. In addition, we reconstructed the $\Sigma$--$D$ probability density function (PDF) using bootstrap based kernel density smoothing \citep[for details see][and references therein]{bozzetto2017}. The reconstruction of the $\Sigma$--$D$ PDF provides a better insight into observational biases and evolutionary features of each sample \citep[see also][and references therein]{Vukotic2014}. Unlike \citet{Vukotic2014}, the PDF calculation is improved with the use of bootstrap based kernel density smoothing instead of bootstrap based centroid offset smoothing. We applied a standard Gaussian kernel in order to calculate optimal smoothing bandwidths. For data in two dimensions a Gaussian product kernel was used. The optimal smoothing bandwidths ($h$) were selected by minimising the bootstrap integrated mean squared error (BIMSE) \cite[for details see][]{bozzetto2017}. 

From the LMC and SMC data we defined three samples: 1) the sample of 28 LMC PNe; 2) the sample of 9 SMC PNe with reliable flux density estimates; 3) the sample consisting of the combined LMC + SMC PNe with 37 total PNe.

\begin{figure*}
\includegraphics[width=0.45\textwidth]{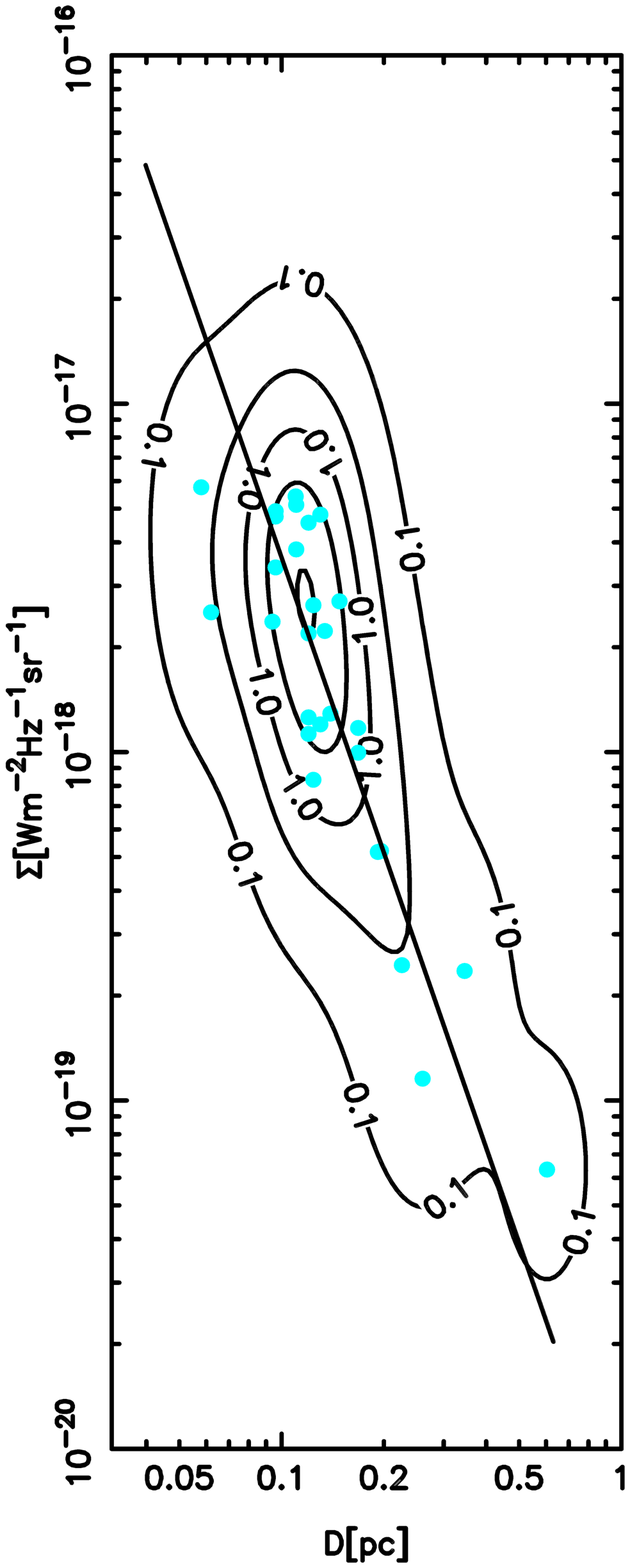}
\includegraphics[width=0.45\textwidth]{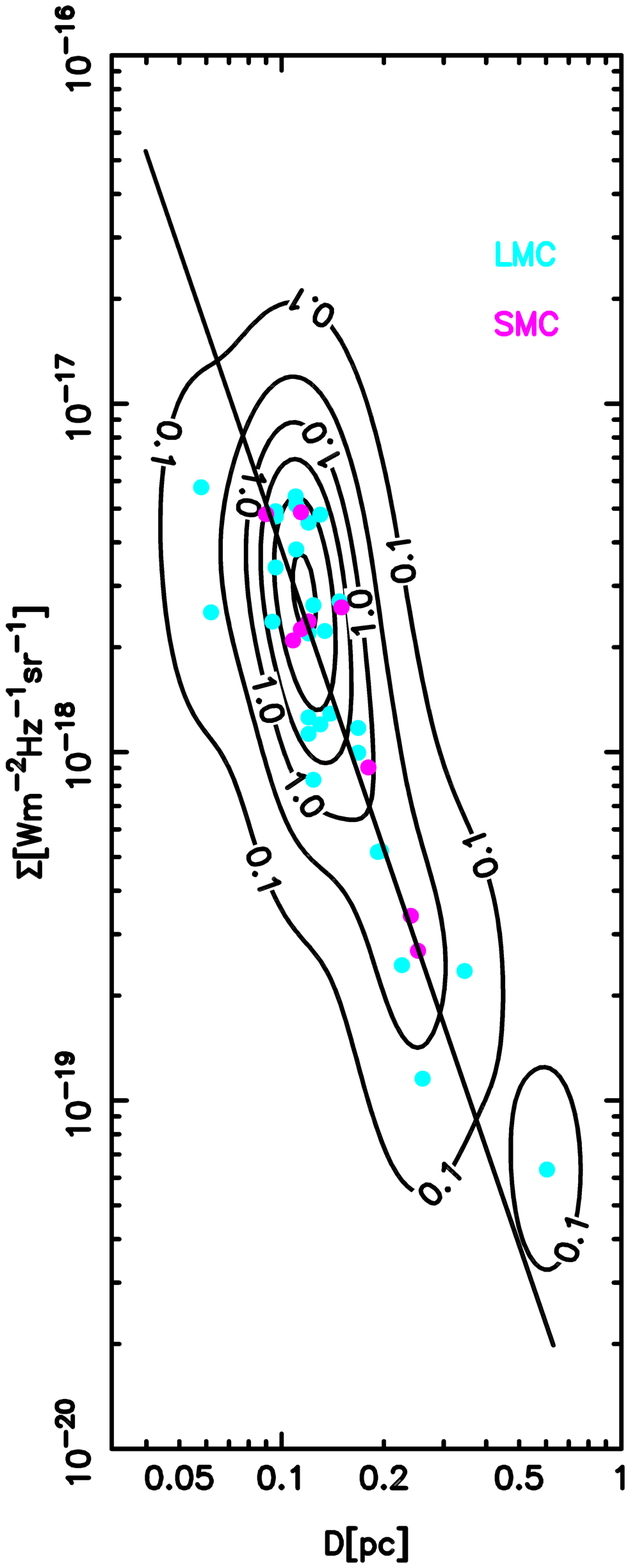}
\caption{$\Sigma$--$D$ data plots at 6~cm with the reconstructed data probability density distribution and orthogonal regression best-fitting line: LMC (left) and LMC+SMC (right). The PDF contour plot is calculated on the $100 \times 100$ grid mapped on the plotted variables range. Contours levels are at 0.1, 0.5, 1.0, 1.5, 2.0 and 2.5. Optimal smoothing bandwidths are found at $h_{\log D} = 0.11$ and $h_{\log \Sigma} = 0.30$ for the LMC sample and $h_{\log D} = 0.09$ and $h_{\log \Sigma} = 0.26$ for the LMC+SMC sample. }
\label{SigmaD_results_figure}
\end{figure*}

As a measure of the sample spread and reliability we calculated the correlation coefficient $r$ and the average fractional error $f$. For the $\Sigma$--$D$ statistical distances to the MCs, we calculated $f$ and estimated the distances from the $\Sigma$--$D$ relation (for a detailed description of statistical distance calculation see \citet{Vukotic2014}). The results are shown in Table~\ref{SigmaD_results_table} and Fig.~\ref{SigmaD_results_figure}.

The average fractional error for distance was calculated as:
\begin{equation}
{f}=\frac{1}{N} \sum_{i=1}^N \left| \frac{d_\mathrm{i}-d_\mathrm{i}^\mathrm{s}}{d_\mathrm{i}} \right|
\end{equation}
\noindent where $N$ is the number of data points in the sample, $d_\mathrm{i}$ is the measured distance to the object of the $\mathrm{i}^\mathrm{th}$ data point, and $d_\mathrm{i}^\mathrm{s}$ is a statistical distance to that object determined either from the parameters of an empirical fitting or from the PDF method. The PDF method for a fixed $\Sigma$ value gives a probability density over $D$. The values of $D$ for the mean, mode, and median are used to calculate the corresponding statistical distances (Table~\ref{SigmaD_results_table}).

\subsection{SMC sample}
With the highest $r$ value and the smallest values for $f$ (Table~\ref{SigmaD_results_table}) the SMC sample appears to be very compact and reliable. However, the kernel density smoothing does not give convergent results for optimal smoothing bandwidths. This is caused by the small number of data points. The optimal smoothing bandwidths are calculated around $\sim 0.14$ and $\sim 0.40$ but repeated calculations do not give consistent results. 

We note that the SMC sample consists of a small number of data points that are difficult to measure accurately in that they are very close to the survey sensitivity line. This suggests that the quality of this sample could be weaker than it would first appear from a simple linear regression. Applying the bootstrap procedure could provide erroneous values for $A$ and $\beta$ and their uncertainties. The small number of data points, even with the high correlation value, gives non-physical values for $A$ and $\beta$ when fitting the re-sampled samples. This resampling with repetition is the essence of the bootstrap analysis (see \citet{Vukotic2014} and references therein).

Instead of presenting these erroneous values, in Table~\ref{SigmaD_results_table} we show values from a simple empirical fitting of a single line to the SMC sample without the application of the bootstrap analysis. Even though we do not consider these fitted parameter values reliable we do utilise the SMC sample to form a more complete combined MC PN sample.

\subsection{LMC and combined sample}
The larger LMC sample appears to be more complete and reliable than the SMC sample although it is also very close to the survey sensitivity line (Figs.~\ref{SigmaD_results_figure} and \ref{sensitivity_figure}). An interesting feature of these plots is that the data density contours in the $\Sigma-D$ plane at $\gtrsim 1$ levels are symmetric about the line of the steeper slope than the slope of the orthogonal offset best-fitting line for both LMC and the combined sample. This (a)symmetric discrepancy is evident over a large part of the plotted variables range, $D\lesssim0.3$ pc. This is in agreement with the fig.~5 plot from \citet{Frew2016} where the authors plotted similar quantities (\Halpha\ brightness verses radius) along with the evolutionary paths for PNe of different masses. The plotted paths significantly deviate from a straight line in the $D\lesssim0.3$~pc range. This reveals the need for developing new statistical tools for future empirical studies of PN evolution in order to quantify the subtleties of that evolution. With the larger data spread, the LMC sample is certainly less sensitivity biased than the SMC sample but the somewhat large uncertainties, especially $\Delta \beta \sim 0.4-0.5 $, indicate a possible reduced sample quality. Although less extreme than the SMC sample, the LMC sample bias could cause peculiar values for the fitting of the parameters for the re-sampled sample. 

Combining the SMC and LMC samples into one sample yielded the best results. The combined sample has smaller values for $\Delta A$ and $\Delta \beta$ compared to the LMC sample alone (Table~\ref{SigmaD_results_table}). We also found that the average fractional distance error calculated with the PDF method for the combined sample is $f\sim16$ per cent which is lower than that of the LMC sample alone which is $f\sim18$ per cent. This demonstrates the potential of RC MC PN samples to be used for statistical distance estimates. However, $\sim18$ per cent of the LMC's $50$~kpc distance is $\sim$9~kpc. Comparing this to a 1~kpc LMC depth \citep{Pietrzynski2013}, \citep{Frew2016b} suggests that there is still significant room for improvement. This is likely to be achieved by more sensitive and more accurate radio surveys with larger samples.

\subsection{$\Sigma$--$D$ slope}
The value of $\beta$ in the $\Sigma$--$D$ relation is theoretically expected to change from $\sim$1 to 3 over the lifetimes of the PNe \citep{Urosevic2009}. Using a GPN sample they found an empirical slope of $\beta \simeq 2.3 \pm 0.4$ by fitting the $\Sigma$ offsets. The $\beta$ values from $\Sigma$ offsets in this work, $2.3$ for combined sample and $2.2$ for the LMC sample alone, may imply a similar age for the MC PN population compared to their selected GPN sample. However, the more robust orthogonal slope of $2.9$ (Table~\ref{SigmaD_results_table}) could give the impression that the MC PNe might be in a late stage of their evolution. \citet{Vukotic2009} argued that the MC PN samples are biased with sensitivity selection effects and consequently the empirical slopes should be steeper than that estimated from the $\Sigma-D$ empirical data fittings. Applying the bootstrap procedure to a sample of GPNe with reliable distances, \citet{2012IAUS..283..522V} obtained an orthogonal offset $\beta = 3.1 \pm 0.4$ which is consistent with our MC results presented in Table~\ref{SigmaD_results_table}. Both the $\Sigma$ offset slope for the GPN sample from \citet{Urosevic2009} and the orthogonal offset slope from \citet{2012IAUS..283..522V} overlap within the estimated uncertainties of the slopes from Table~\ref{SigmaD_results_table}.

\subsection{$\Sigma$--$D$ sensitivity selection effects}
\label{sensitivity}
To estimate the accuracy of our results related to the $\Sigma$--$D$ slope we simulated the influence of sensitivity related selection effects. This method is from \cite{Vukotic2009} who applied this technique to a very small sample of five PNe. We performed a similar simulation (although more advanced in some technical details) applying this method to our MC PN samples. This great improvement in the number of radio MC PN detections should result in a significantly more complete PN sample.

\begin{figure}
 \includegraphics[width=0.475\textwidth]{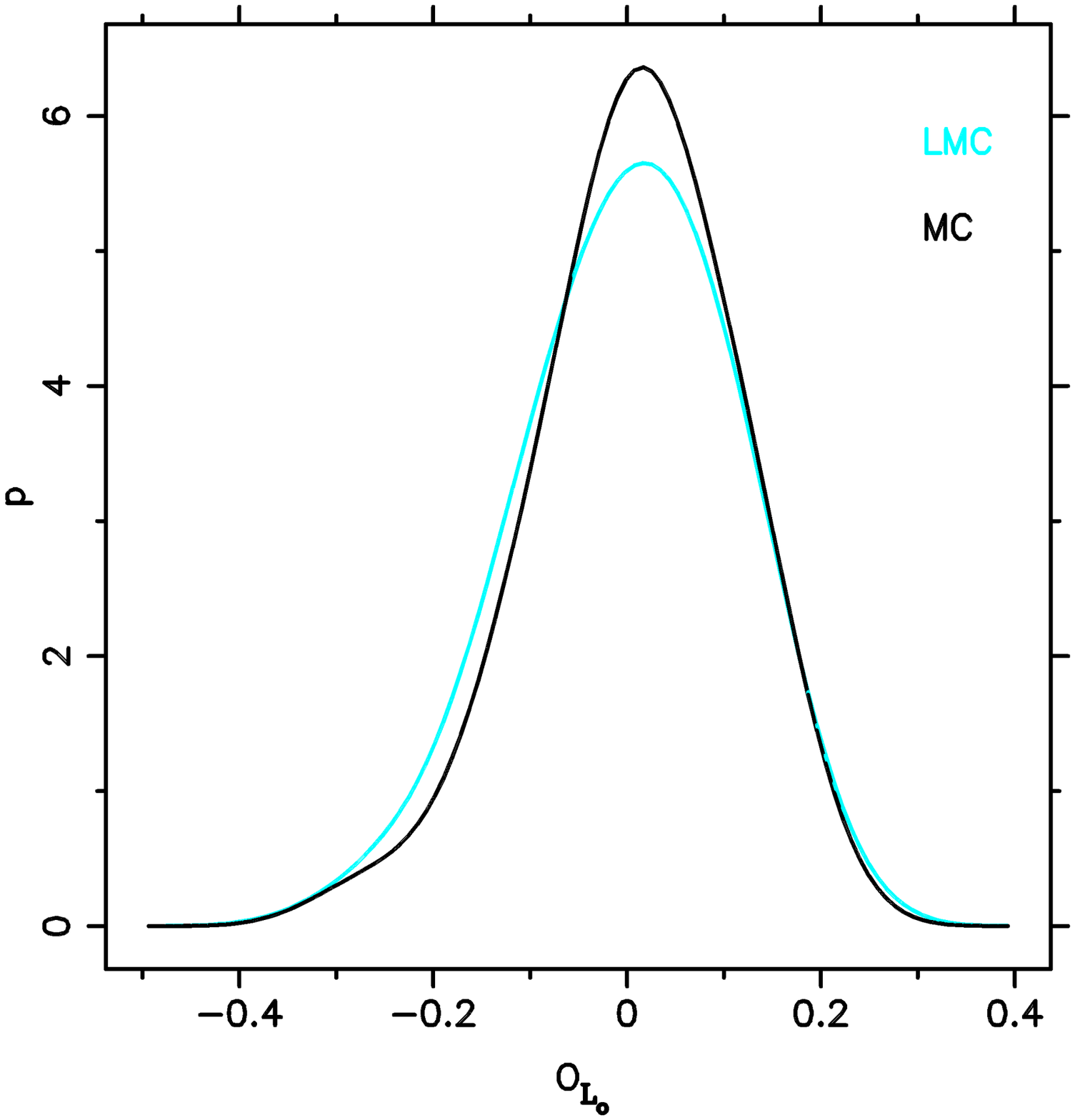}
 \caption{Our initial findings for the 6~cm data sample probability density distributions of the orthogonal distances from the orthogonal offset best-fitting lines $L_\mathrm{o}$ for the LMC and the combined sample.}
 \label{sensitivity_pdf}
\end{figure}

To estimate the influence of the sensitivity selection effects we generated artificial PN $\Sigma$--$D$ samples using a Monte Carlo simulation from the properties of our 6~cm data and compared the sample parameters before and after sensitivity selection. With the smallest PN in our 6~cm sample having a diameter of $D_\mathrm{min}=0.058$~pc (Table~\ref{tbl:physprop}) and the fact that the largest PNe usually swell up to $D_\mathrm{max}\sim2$~pc, we distributed artificial points randomly generated in the $[D_\mathrm{min}, D_\mathrm{max}]$ interval using the random number generator of \cite{saito_matsumoto08} with a uniform density on the $\log D$ scale equal to the density of the adopted 6~cm data. This resulted in a total of 55 artificial points (for the combined sample and 42 for the LMC sample) per generated sample in a given $[\log D_\mathrm{min}, \log D_\mathrm{max}]$ interval. These points were then projected on to a series of lines with $\beta$ ranging from $1.5$ to $3.6$. All of these lines intersected the orthogonal offset best-fitting line to the 6~cm adopted data ($L_\mathrm{o}$) at $D_\mathrm{min}$. The next step was to disperse the points from the lines upon which they were projected in such a way as to simulate the dispersion of the adopted 6~cm data. This was achieved by applying the kernel density smoothing in 1D from \citet{bozzetto2017}. From the adopted 6~cm data we reconstructed the smooth probability density distribution of the orthogonal distances (offsets) from the $L_\mathrm{o}$ line (Fig.~\ref{sensitivity_pdf}).

\begin{figure}
\includegraphics[width = 0.47\textwidth]{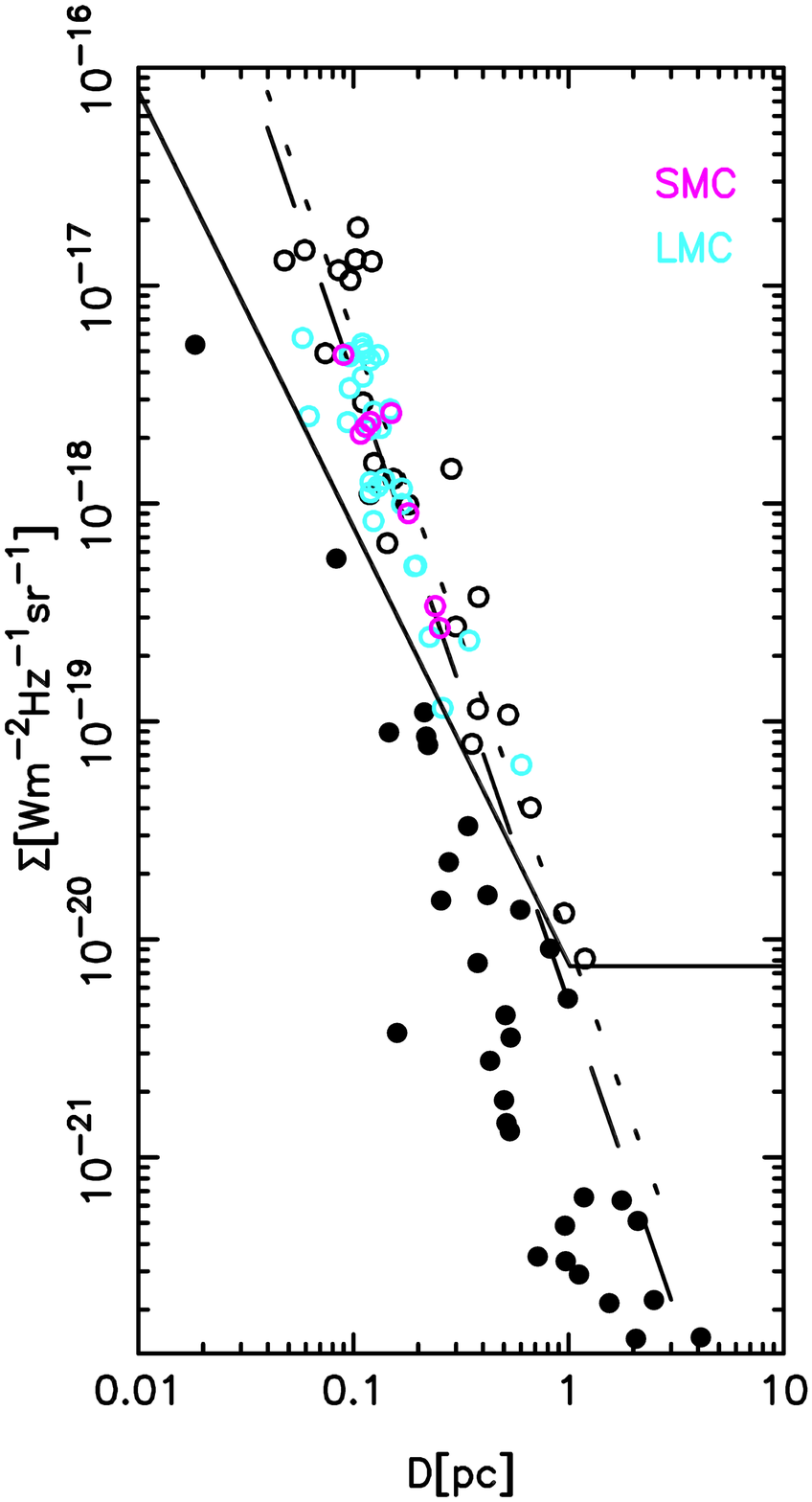}
\caption{ An example of an artificial sample generated from the combined sample orthogonal offsets for the simulated $\beta$ of $3.5$ and a $Scatter$ of 2.0. After selection the slope flattens to $2.8$ (short-dashed line fit). The solid line is the sensitivity line with the horizontal part at $\Sigma = 7.54\times10^{-21}$Wm$^{-2}$Hz$^{-1}$sr$^{-1}$ with the break point at $D=1.017$~pc. The long-dashed line is the orthogonal offset best-fitting to the combined MC PN sample. Data points in black are the simulated sample with the points below the sensitivity line plotted as filled circles.}
\label{sensitivity_figure}
\end{figure}

It is reasonable to assume that the data do not have the same scatter along the simulated interval. We used empirical data for the input of our sensitivity simulation and we modeled the sensitivity line in such a way that it passed through the data point that gave the lowest $\Sigma$ value. We have no empirical knowledge of the data spread at the faint end. Even for the part above the sensitivity line, the PDF contours from Fig.~\ref{SigmaD_results_figure} do not appear to have much variability in spread. In this way we simulated the uniform dispersion model for the whole simulated data interval. This is sufficient to demonstrate the influence of the sensitivity selection on the $\Sigma$--$D$ slope. Even if we use a different \textit{Scatter} value at the faint end, most of it will get a sensitivity cut off and will not influence the results. Our initial findings (Fig.~\ref{sensitivity_pdf}) showed no significant difference for the overall scatter in the case of the LMC and the combined sample. Both distributions were rather symmetric around the peak which was close to a value of zero. Each point was randomly dispersed in the orthogonal direction from the simulated slope line according to the probability density distribution from Fig.~\ref{sensitivity_pdf}. The dispersion was controlled with a parameter called $Scatter$ which is multiplied by the dispersion obtained from the distributions in Fig.~\ref{sensitivity_pdf} to obtain the actual dispersion. We then calculated the orthogonal offset fitting parameters for the dispersed artificial sample. For each simulated slope we generated $100$ artificial samples. The mean values of the fitting parameters after selection and their estimated uncertainties (calculated as the standard deviation of the given array of elements) are presented in Table~\ref{sensitivity_table}. In Fig.~\ref{sensitivity_figure} we plotted one artificial sample for $Scatter=2.0$ with a simulated $\beta$ of $3.5$ along with the combined MC PN sample and sensitivity line. Since the presented LMC survey is not uniform in terms of sensitivity we selected the sensitivity line in such a way that it passed through the data point of the combined sample that gives the lowest $\Sigma$ value at the horizontal part of the sensitivity line. For the break point of the sensitivity line we selected the average equivalent of the circular beam size calculated from the beam sizes of each particular reduced image. At the adopted LMC distance this translates to $D=1.017$~pc. 

\begin{table*}
\caption{Parameters of the simulated samples and results of the sensitivity simulations for the LMC sample and the combined sample. Cols. 1-2 are the parameters from Equation \ref{SigD_equation} of the simulated $\Sigma$--$D$ evolution lines. While Cols.~3-7 present the results for LMC sample, Cols.~8-12 in a same way present the results for the combined sample. In Cols.~3 and 5 are the mean values of $\Sigma$--$D$ parameters calculated from the orthogonal offset fits after the selection has been applied and their estimated uncertainties are in Cols. ~4 and 6, respectfully. In Col.~7 we give the difference between the mean slope after selection and the simulated slope. For all simulated slopes we fit 100 artificially generated samples. }
\scriptsize
\begin{tabular}{cccccccrcccccr}\hline
& & & \multicolumn{5}{c}{LMC sample}&&\multicolumn{5}{c}{Combined sample}\\\cline{4-8} \cline{10-14}
\multicolumn{2}{c}{Simulated}&&\multicolumn{4}{c}{After selection}&&&\multicolumn{4}{c}{After selection}&\\\cline{1-2}\cline{4-7}\cline{10-13}
$\log A$ & $\beta$ && $<\log A>$ & $\sigma_{<\log A>}$ & $<\beta>$ & $\sigma_{<\beta>}$ & $\Delta\beta$  &&$<\log A>$ & $\sigma_{<\log A>}$ & $<\beta>$ & $\sigma_{<\beta>}$ & $\Delta\beta$  \\
(1) & (2) && (3) & (4) & (5) & (6) & (7) && (8)& (9) & (10) & (11) & (12) \\\hline\hline
\multicolumn{14}{l}{$Scatter$ =  1.0}\\
-18.62 & 1.50 && -18.63 &  0.049 & 1.49 & 0.080 &  0.01  && -18.56 &  0.035 & 1.48 & 0.054 &  0.02 \\ 
-18.74 & 1.60 && -18.75 &  0.054 & 1.59 & 0.083 &  0.01  && -18.70 &  0.033 & 1.59 & 0.053 &  0.01 \\ 
-18.87 & 1.70 && -18.87 &  0.053 & 1.68 & 0.077 &  0.02  && -18.82 &  0.037 & 1.69 & 0.060 &  0.01 \\ 
-18.99 & 1.80 && -19.00 &  0.057 & 1.79 & 0.093 &  0.01  && -18.94 &  0.040 & 1.79 & 0.065 &  0.01 \\ 
-19.11 & 1.90 && -19.12 &  0.054 & 1.90 & 0.089 &  0.00  && -19.07 &  0.045 & 1.89 & 0.062 &  0.01 \\ 
-19.24 & 2.00 && -19.24 &  0.070 & 2.00 & 0.112 &  0.00  && -19.19 &  0.048 & 1.98 & 0.069 &  0.02 \\ 
-19.36 & 2.10 && -19.36 &  0.057 & 2.09 & 0.098 &  0.01  && -19.31 &  0.049 & 2.09 & 0.075 &  0.01 \\ 
-19.48 & 2.20 && -19.48 &  0.064 & 2.20 & 0.107 & -0.00  && -19.44 &  0.048 & 2.19 & 0.071 &  0.01 \\ 
-19.61 & 2.30 && -19.59 &  0.061 & 2.26 & 0.098 &  0.04  && -19.55 &  0.051 & 2.27 & 0.076 &  0.03 \\ 
-19.73 & 2.40 && -19.68 &  0.065 & 2.35 & 0.116 &  0.05  && -19.66 &  0.049 & 2.36 & 0.084 &  0.04 \\ 
-19.86 & 2.50 && -19.78 &  0.078 & 2.40 & 0.120 &  0.10  && -19.76 &  0.064 & 2.44 & 0.099 &  0.06 \\ 
-19.98 & 2.60 && -19.86 &  0.077 & 2.46 & 0.120 &  0.14  && -19.85 &  0.063 & 2.50 & 0.102 &  0.10 \\ 
-20.10 & 2.70 && -19.94 &  0.097 & 2.53 & 0.159 &  0.17  && -19.93 &  0.066 & 2.56 & 0.107 &  0.14 \\ 
-20.23 & 2.80 && -20.00 &  0.102 & 2.53 & 0.152 &  0.27  && -20.01 &  0.075 & 2.61 & 0.110 &  0.19 \\ 
-20.35 & 2.90 && -20.07 &  0.136 & 2.61 & 0.167 &  0.29  && -20.09 &  0.089 & 2.68 & 0.127 &  0.22 \\ 
-20.47 & 3.00 && -20.17 &  0.163 & 2.71 & 0.227 &  0.29  && -20.16 &  0.099 & 2.72 & 0.132 &  0.28 \\ 
-20.60 & 3.10 && -20.18 &  0.144 & 2.71 & 0.197 &  0.39  && -20.22 &  0.137 & 2.78 & 0.176 &  0.32 \\ 
-20.72 & 3.20 && -20.27 &  0.183 & 2.77 & 0.244 &  0.43  && -20.28 &  0.128 & 2.81 & 0.168 &  0.39 \\ 
-20.85 & 3.30 && -20.32 &  0.262 & 2.79 & 0.323 &  0.51  && -20.39 &  0.183 & 2.91 & 0.219 &  0.39 \\ 
-20.97 & 3.40 && -20.37 &  0.224 & 2.84 & 0.277 &  0.56  && -20.45 &  0.169 & 2.96 & 0.209 &  0.44 \\ 
-21.09 & 3.50 && -20.46 &  0.433 & 2.92 & 0.444 &  0.58  && -20.58 &  0.294 & 3.05 & 0.338 &  0.45 \\ 
-21.22 & 3.60 && -20.55 &  0.392 & 3.00 & 0.413 &  0.60  && -20.61 &  0.281 & 3.08 & 0.308 &  0.52 \\ 
\multicolumn{14}{l}{$Scatter$ =  2.0}\\
-18.62 & 1.50 && -18.62 &  0.091 & 1.50 & 0.144 & -0.00  && -18.56 &  0.069 & 1.48 & 0.115 &  0.02 \\ 
-18.74 & 1.60 && -18.73 &  0.099 & 1.59 & 0.145 &  0.01  && -18.70 &  0.074 & 1.60 & 0.123 &  0.00 \\ 
-18.87 & 1.70 && -18.87 &  0.107 & 1.71 & 0.182 & -0.01  && -18.80 &  0.077 & 1.68 & 0.115 &  0.02 \\ 
-18.99 & 1.80 && -18.96 &  0.095 & 1.80 & 0.153 & -0.00  && -18.92 &  0.065 & 1.77 & 0.105 &  0.03 \\ 
-19.11 & 1.90 && -19.09 &  0.101 & 1.90 & 0.165 &  0.00  && -19.05 &  0.082 & 1.88 & 0.131 &  0.02 \\ 
-19.24 & 2.00 && -19.17 &  0.099 & 1.97 & 0.146 &  0.03  && -19.17 &  0.083 & 1.99 & 0.125 &  0.01 \\ 
-19.36 & 2.10 && -19.29 &  0.100 & 2.09 & 0.151 &  0.01  && -19.27 &  0.075 & 2.08 & 0.136 &  0.02 \\ 
-19.48 & 2.20 && -19.34 &  0.116 & 2.10 & 0.198 &  0.10  && -19.36 &  0.084 & 2.14 & 0.141 &  0.06 \\ 
-19.61 & 2.30 && -19.42 &  0.111 & 2.17 & 0.173 &  0.13  && -19.43 &  0.073 & 2.18 & 0.122 &  0.12 \\ 
-19.73 & 2.40 && -19.48 &  0.137 & 2.20 & 0.215 &  0.20  && -19.52 &  0.078 & 2.26 & 0.129 &  0.14 \\ 
-19.86 & 2.50 && -19.56 &  0.129 & 2.29 & 0.212 &  0.21  && -19.59 &  0.101 & 2.33 & 0.174 &  0.17 \\ 
-19.98 & 2.60 && -19.62 &  0.140 & 2.30 & 0.223 &  0.30  && -19.66 &  0.106 & 2.36 & 0.182 &  0.24 \\ 
-20.10 & 2.70 && -19.68 &  0.145 & 2.37 & 0.225 &  0.33  && -19.72 &  0.110 & 2.42 & 0.184 &  0.28 \\ 
-20.23 & 2.80 && -19.74 &  0.166 & 2.44 & 0.237 &  0.36  && -19.80 &  0.129 & 2.48 & 0.196 &  0.32 \\ 
-20.35 & 2.90 && -19.77 &  0.156 & 2.46 & 0.259 &  0.44  && -19.84 &  0.142 & 2.50 & 0.210 &  0.40 \\ 
-20.47 & 3.00 && -19.79 &  0.184 & 2.45 & 0.321 &  0.55  && -19.88 &  0.169 & 2.53 & 0.230 &  0.47 \\ 
-20.60 & 3.10 && -19.85 &  0.198 & 2.48 & 0.278 &  0.62  && -19.95 &  0.141 & 2.60 & 0.193 &  0.50 \\ 
-20.72 & 3.20 && -19.90 &  0.216 & 2.54 & 0.311 &  0.66  && -19.97 &  0.180 & 2.58 & 0.252 &  0.62 \\ 
-20.85 & 3.30 && -19.95 &  0.227 & 2.59 & 0.329 &  0.71  && -20.02 &  0.226 & 2.62 & 0.308 &  0.68 \\ 
-20.97 & 3.40 && -20.02 &  0.302 & 2.67 & 0.440 &  0.73  && -20.01 &  0.212 & 2.62 & 0.260 &  0.78 \\ 
-21.09 & 3.50 && -20.09 &  0.397 & 2.72 & 0.541 &  0.78  && -20.10 &  0.290 & 2.71 & 0.388 &  0.79 \\ 
-21.22 & 3.60 && -20.06 &  0.343 & 2.65 & 0.479 &  0.95  && -20.18 &  0.290 & 2.79 & 0.346 &  0.81 \\ 
\multicolumn{14}{l}{$Scatter$ =  3.0}\\
-18.62 & 1.50 && -18.54 &  0.127 & 1.49 & 0.233 &  0.01  && -18.51 &  0.091 & 1.47 & 0.167 &  0.03 \\ 
-18.74 & 1.60 && -18.67 &  0.129 & 1.59 & 0.217 &  0.01  && -18.66 &  0.119 & 1.60 & 0.187 & -0.00 \\ 
-18.87 & 1.70 && -18.78 &  0.135 & 1.72 & 0.243 & -0.02  && -18.76 &  0.104 & 1.70 & 0.173 & -0.00 \\ 
-18.99 & 1.80 && -18.86 &  0.149 & 1.78 & 0.254 &  0.02  && -18.88 &  0.105 & 1.81 & 0.182 & -0.01 \\ 
-19.11 & 1.90 && -18.98 &  0.137 & 1.92 & 0.233 & -0.02  && -18.96 &  0.119 & 1.87 & 0.209 &  0.03 \\ 
-19.24 & 2.00 && -19.07 &  0.128 & 1.99 & 0.206 &  0.01  && -19.06 &  0.105 & 1.97 & 0.169 &  0.03 \\ 
-19.36 & 2.10 && -19.14 &  0.158 & 2.01 & 0.298 &  0.09  && -19.18 &  0.122 & 2.10 & 0.176 & -0.00 \\ 
-19.48 & 2.20 && -19.18 &  0.130 & 2.07 & 0.191 &  0.13  && -19.25 &  0.102 & 2.14 & 0.170 &  0.06 \\ 
-19.61 & 2.30 && -19.22 &  0.137 & 2.08 & 0.230 &  0.22  && -19.31 &  0.116 & 2.19 & 0.208 &  0.11 \\ 
-19.73 & 2.40 && -19.30 &  0.153 & 2.21 & 0.287 &  0.19  && -19.36 &  0.128 & 2.25 & 0.199 &  0.15 \\ 
-19.86 & 2.50 && -19.35 &  0.165 & 2.25 & 0.280 &  0.25  && -19.43 &  0.130 & 2.24 & 0.214 &  0.26 \\ 
-19.98 & 2.60 && -19.37 &  0.159 & 2.26 & 0.338 &  0.34  && -19.47 &  0.131 & 2.28 & 0.249 &  0.32 \\ 
-20.10 & 2.70 && -19.44 &  0.198 & 2.35 & 0.378 &  0.35  && -19.53 &  0.132 & 2.35 & 0.210 &  0.35 \\ 
-20.23 & 2.80 && -19.51 &  0.190 & 2.35 & 0.357 &  0.45  && -19.56 &  0.145 & 2.36 & 0.234 &  0.44 \\ 
-20.35 & 2.90 && -19.55 &  0.221 & 2.44 & 0.394 &  0.46  && -19.58 &  0.156 & 2.37 & 0.246 &  0.53 \\ 
-20.47 & 3.00 && -19.57 &  0.266 & 2.41 & 0.448 &  0.59  && -19.66 &  0.184 & 2.47 & 0.273 &  0.53 \\ 
-20.60 & 3.10 && -19.57 &  0.251 & 2.40 & 0.438 &  0.70  && -19.70 &  0.166 & 2.49 & 0.286 &  0.61 \\ 
-20.72 & 3.20 && -19.69 &  0.266 & 2.58 & 0.438 &  0.62  && -19.72 &  0.204 & 2.51 & 0.317 &  0.69 \\ 
-20.85 & 3.30 && -19.66 &  0.244 & 2.46 & 0.433 &  0.84  && -19.78 &  0.283 & 2.58 & 0.410 &  0.72 \\ 
-20.97 & 3.40 && -19.71 &  0.310 & 2.60 & 0.529 &  0.80  && -19.78 &  0.203 & 2.52 & 0.328 &  0.88 \\ 
-21.09 & 3.50 && -19.73 &  0.275 & 2.59 & 0.499 &  0.91  && -19.83 &  0.250 & 2.58 & 0.382 &  0.92 \\ 
-21.22 & 3.60 && -19.76 &  0.313 & 2.60 & 0.493 &  1.00  && -19.87 &  0.326 & 2.63 & 0.504 &  0.97 \\ 
\hline
 \end{tabular}
 \label{sensitivity_table}
\end{table*}

The results in Table~\ref{sensitivity_table} do not appear to be significantly dependent on $Scatter$. For each given $Scatter$ the $\Delta\beta$ becomes larger than $\sigma_{<\beta>}$ when the simulated $\beta$ becomes as large as $\sim 2.5-2.7$. This corresponds to $<\beta>\sim 2.5$ ($Scatter=1.0$) and, although within the estimated uncertainty, reduces to $<\beta> \sim2.3$ for a larger $Scatter$. Hence the steeper slopes, such as $\sim 2.9$ (as obtained in this work) are significantly influenced by the sensitivity selection effect. For $<\beta> \sim 2.9$ the simulated slope is $\sim 3.4$ and is even greater for larger $Scatter$. Since the MC PN samples appear to be significantly influenced by sensitivity selection it is likely that only the brightest part of the sample is detected. However, any strong claims should be avoided since the results might depend on the simulation algorithm and the selected sensitivity line. 

\section {Summary} 
 \label{s:Summary}
 
We searched the ATOA for projects with high resolution data at 3, 6, 13, or 20~cm which had pointing centres within 10~arcmin of PN coordinates from the RP database. Some seventeen projects met those requirements and were processed in order to search for PNe at established (RP database) coordinates. We found 28 sources with detectable radio emissions in at least one of these radio bands with a flux density $\geq3\sigma$ above the noise. Of these 28 PNe, 21 are new radio detections reported here for the first time. We also report a total of 61 ATCA images from the seventeen projects within which we did not find any PN RC emissions at RP database coordinates that were $\geq3\sigma $ above the noise.

This radio-continuum tally is only $\sim$5 per cent of the known population of PNe in the LMC. The likely reasons for this very small detection ratio are the relatively low flux densities from the PNe at the distance to the LMC and the lack of complete  (and uniform) area coverage of the LMC with high sensitivity observations.

Our measurements along with data from the literature were used to calculate additional RC properties of the detected PNe. We calculated the MIR-20~cm flux density ratios resulting in a median value of 11.9 compared to the \cite{Cohen2007} value for the Galactic bulge of 12. We further calculated essentially no self absorption effect for our 6~cm measurements. In addition, our adopted 6~cm flux densities and diameters compare very well with the Galactic bulge distance-scaled PN data set from \cite{Siodmiak2001}.

We used the combined sample of $37$ radio PNe from the LMC and the SMC \citep{leverenz2016} with available reliable data to examine the $6$~cm radio $\Sigma$--$D$ relation. Two sample sets were analysed: the LMC sample and the combined LMC+SMC sample. The reconstruction of the probability density function shows that the rather complex PN evolutionary paths are not likely to be well represented by a single linear best-fitting line but that more complex statistical tools should be used. The rather incomplete SMC sample well complements the more numerous LMC sample. We fitted the selected $\Sigma$--$D$ data with both $\Sigma$ offsets and orthogonal offsets to obtain the fitting parameters. The value of $\beta$ in the $\Sigma$--$D$ relation is theoretically expected to change from $\sim$1 to 3 over the lifetime of the PNe \citep{Urosevic2009}. For the combined sample we obtained for orthogonal offsets $\beta=2.9\pm0.4$ and for $\Sigma$ offsets $\beta=2.3\pm0.3$. Both values are comparable to the previously obtained values for GPN samples. 

Sensitivity selection effects were examined by simulating an artificial random sample with values of $\beta$ from $1.5$ to $3.6$ and values of $D$ extending over a range from the minimum observed diameter in our sample to $2$~pc. A strong selection effect is seen in MC data. The MC RC data is very close to the measurement sensitivity limit which weakens its quality and $\Sigma$ is only detected in the high range of values expected from measurements taken of GPNe. The results of our Monte Carlo sensitivity simulation suggested that selection effects are significant for values larger than $\beta \sim 2.6$. We note that a measured slope of $\beta=2.9$ should correspond to a sensitivity free value of $\sim 3.4$. We demonstrated that an artificial sample derived from the combined sample shows that a large part of the data contained in the artificial sample cannot be detected with the sensitivity of the measurements in available RC data from the MCs.

For the LMC and combined LMC+SMC samples, applying the PDF statistical distance calculation, we calculated the average fractional error for the statistical distance measurement. For the mean, mode, and median PDF estimators the average fractional distance error was $\sim16$ per cent for the combined sample and $\sim18$ per cent for the LMC sample alone. At the LMC distance ($\sim 50$~kpc) this $\sim18$ per cent translates to $\sim 9$~kpc average distance error which is still significantly larger than the $1$~kpc LMC depth. 

Our study demonstrated the potential for  extended RC MC PN samples to be used as statistical distance estimators. Still, much can be improved, including the understanding of observational biases, especially sensitivity selection effects. Next generation multi-frequency radio surveys with greater sensitivity  and accuracy should lead to improved  PN classification and  to a better understanding of  PDF evolutionary features.

\section*{Acknowledgements}

The Australia Telescope Compact Array is part of the Australia Telescope National Facility which is funded by the Commonwealth of Australia for operation as a National Facility managed by CSIRO. This paper includes archived data obtained through the Australia Telescope Online Archive (http://atoa.atnf.csiro.au). This research has made use of Aladin, SIMBAD and VizieR, operated at the CDS, Strasbourg, France. We used the {\sc karma} and {\sc miriad} software packages developed by the ATNF and the {\sc python} programming language. DU and BV acknowledge financial support from the Ministry of Education, Science and Technological Development of the Republic of Serbia through the project \#176005 `Emission nebulae: structure and evolution'. We would like to thank the anonymous referee for the astute comments and careful reading of this paper that significantly improved the final version.

\bibliographystyle{mnras}


\bibliography{leverenz_bib}                


\bsp	
\label{lastpage}
\end{document}